%% file: EXO-10-017_temp.tex
\pdfoutput=1
\documentclass[11pt,twoside,a4paper,cmspaper,final,collab]{cms-tdr}
\def\svnVersion{26811M}\def\cmsCernNo{2010-073}\def\cmsCernDate{\today}\def\cmsMessage{Submitted to Physics Letters B}

\begin{document}\cmsNoteHeader{EXO-10-017}
\hyphenation{env-iron-men-tal}
\hyphenation{had-ron-i-za-tion}
\hyphenation{cal-or-i-me-ter}
\hyphenation{de-vices}
\RCS$Revision: 26802 $
\RCS$HeadURL: svn+ssh://alverson@svn.cern.ch/reps/tdr2/papers/EXO-10-017/trunk/EXO-10-017.tex $
\RCS$Id: EXO-10-017.tex 26802 2010-12-14 23:53:36Z glandsbe $
\input{ptdr-definitions}

\def\MP{\mbox{$M_D$}}
\def\mp{\mbox{$M_D$}\ }
\def\TH{\mbox{$T_H$}\ }
\def\mbh{\mbox{$M_{\rm BH}$}\ }     
\def\MBH{\mbox{$M_{\rm BH}$}}         
\def\MET{\mbox{${\hbox{$E$\kern-0.6em\lower-.1ex\hbox{/}}}_T$}} 
\def\met{\mbox{${\hbox{$E$\kern-0.6em\lower-.1ex\hbox{/}}}_T$}\ } 
\def\ipb{pb$^{-1}$}                     
\def\etal{{\sl et al.}}                 
\def\vs{{\sl vs.}}                      
\def\et{\mbox{$E_T$}}
\cmsNoteHeader{EXO-10-017} 
\title{Search for Microscopic Black Hole Signatures at the Large Hadron Collider}
\address[cern]{CERN}
\author[cern]{The CMS Collaboration}
\date{\today}
\abstract{
A search for microscopic black hole production and decay in $pp$ collisions at a center-of-mass energy of 7 TeV has been conducted by the CMS Collaboration at the LHC, using a data sample corresponding to an integrated luminosity of 35~pb$^{-1}$. Events with large total transverse energy are analyzed for the presence of multiple high-energy jets, leptons, and photons, typical of a signal expected from a microscopic black hole. Good agreement with the expected standard model backgrounds, dominated by QCD multijet production, is observed for various final-state multiplicities. Limits on the minimum black hole mass are set, in the range 3.5 -- 4.5~TeV, for a variety of parameters in a model with large extra dimensions, along with model-independent limits on new physics in these final states. These are the first direct limits on black hole production at a particle accelerator.}
\hypersetup{%
pdfauthor={CMS Collaboration},%
pdftitle={Search for Microscopic Black Hole Signatures at the Large Hadron Collider},%
pdfsubject={CMS},%
pdfkeywords={CMS, physics, black holes, extra dimensions}}

\maketitle 

One of the exciting predictions of theoretical models with extra spatial dimensions and low-scale quantum gravity is the possibility of copious production of microscopic black holes in particle collisions at the CERN Large Hadron Collider (LHC)~\cite{dl,gt}.
Models with low-scale gravity are aimed at solving the hierarchy problem, the puzzlingly large difference between the electroweak and Planck scales.

In this Letter we focus on microscopic black hole production in a model with large, flat, extra spatial dimensions, proposed by Arkani-Hamed, Dimopoulos, and Dvali, and referred to as the ADD model~\cite{add,add1}. This model alleviates the hierarchy problem by introducing $n$ extra dimensions in space, compactified on an $n$-dimensional torus or sphere with radius $r$. The multidimensional space-time is only open to the gravitational interaction, while the gauge interactions are localized on the $3+1$ space-time membrane. As a result, the gravitational coupling is enhanced at distances smaller than $r$, and Newton's law of gravitation is modified at short distances. The ``true'' Planck scale in $4+n$ dimensions ($M_D$) is consequently lowered to the electroweak scale, much smaller than the apparent Planck scale of $M_{\rm Pl} \sim 10^{16}$~TeV seen by a $3+1$ space-time observer. The relationship between $M_D$ and $M_{\rm Pl}$ follows from Gauss's law and is given as $M_{\rm Pl}^2 = 8\pi M_D^{n + 2} r^n$, using the Particle Data Group (PDG) definition~\cite{PDG}.

Such a change in space-time structure and subsequent strengthening of the gravitational field in the ADD model
could allow black hole formation in particle collisions at energies greater than $M_D$, rather than
 $M_{\rm Pl}$, which is the case for a truly 4-dimensional world. Colliding particles would collapse in a black
 hole if their impact parameter were smaller than the Schwarzschild radius of a black hole with the mass
 $M_{\rm BH}$ equal to the total energy accessible in the collision. The Schwarzschild radius of a black hole with
 mass $M_{\rm BH}$ embedded in $4+n$ space-time can be found by solving Einstein's general relativity equations
 and is given by~\cite{mp,adm}:

$$r_S = \frac{1}{\sqrt \pi M_D} {\left[ \frac{M_{\rm BH}}{M_D} \frac{8\Gamma\!\left(\frac{n + 3}{2}\right)}{n + 2} \right]}^{\frac{1}{n + 1}}.
$$

The parton-level cross section of black hole production is derived from
geometrical considerations and is given by $\sigma \sim \pi r_S^2$~\cite{dl,gt}. At LHC energies, this cross section can reach  100~pb for $M_D$ of 1~TeV. The exact cross section cannot be calculated without knowledge of the underlying theory of quantum gravity and is subject to significant uncertainty. It is commonly accepted~\cite{dl,gt} that the minimum black hole mass $M_{\rm BH}^{\rm min}$ cannot be smaller than $M_D$; although the formation threshold can be significantly larger than
this. When a black hole is formed, some fraction of the colliding parton energy may not be trapped within the event horizon and will be emitted in the form of gravitational shock waves, which results in energy, momentum, and angular momentum loss. This effect is particularly model-dependent for black hole masses close to $M_D$. In general, black holes in particle collisions are produced with non-zero angular momentum, which also affects their properties and production cross section.

Once produced, the microscopic black holes would decay thermally via Hawking radiation
\cite{Hawking},
democratically (with
equal probabilities) to all standard model (SM) degrees of freedom. Quarks and gluons are the dominant particles produced in the black hole evaporation ($\sim 75\%$) because they have a large number of color degrees of freedom. The remaining fraction is accounted for by leptons, $W$ and $Z$ bosons, photons, and possibly Higgs bosons. Emission of gravitons by a black hole in the bulk space is generally expected to be suppressed~\cite{EHM}. In some models the evaporation is terminated earlier, when the black hole mass reaches $M_D$, with the formation of a stable non-interacting and non-accreting remnant. The Hawking temperature for a black hole in
$4+n$ space-time is given by~\cite{adm,dl,gt}: $T_H = \frac{n+1}{4\pi r_S}$ (in Planck units $\hbar =c = k_B = 1$, where $k_B$ is the Boltzmann constant) and is typically in the range of a few hundred GeV. The lifetime for such a microscopic black hole is $\sim 10^{-27}$~s~\cite{adm,dl,gt}.

Here we consider semi-classical black holes, whose properties are similar to those for classical black holes described by general relativity and whose mass is close enough to $M_D$ so that quantum effects can not be ignored completely. There are also models~\cite{RM,Calmet,DG} of quantum black holes that decay before they thermalize, mainly into two-jet final states. We do not consider this signature here, leaving it for dedicated searches in the dijet channel~\cite{dijets1,dijets2}.

The microscopic black holes produced at the LHC would be distinguished by high multiplicity, democratic, and highly isotropic decays with the final-state particles carrying hundreds of GeV of energy. Most of these particles would be reconstructed as jets of hadrons. Observation of such spectacular signatures would provide direct information on the
nature of black holes as well as the structure and dimensionality of space-time~\cite{dl}. Microscopic black hole properties are reviewed in more detail in~\cite{review1,review2}.

The search for black holes is based on $\sqrt{s} = 7$~TeV $pp$ collision data recorded by the Compact Muon Solenoid (CMS) detector at the LHC between March and October 2010, which correspond to an integrated luminosity of $34.7 \pm 3.8 \pbinv$. A detailed description of the CMS experiment
can be found elsewhere~\cite{CMS}. The central feature of the CMS detector
is the 3.8~T superconducting solenoid enclosing the
silicon pixel and strip tracker, the electromagnetic
calorimeter (ECAL), and the brass-scintillator hadronic calorimeter
(HCAL). For triggering purposes and to facilitate jet reconstruction, the calorimeter cells are grouped in projective towers, of granularity
$\Delta \eta \times \Delta \phi = 0.087\times0.087$ at central
rapidities and $0.175\times0.175$ in the forward region. Here, the pseudorapidity $\eta$ is defined as $-\ln(\tan\frac{\theta}{2})$, where $\theta$ is the polar angle with respect to the direction of the counterclockwise beam, and $\phi$ is the azimuthal angle.
Muons are measured in the pseudorapidity window $|\eta|< 2.4$ in gaseous detectors embedded in the steel
return yoke.

The CMS trigger system consists of two levels. The first level (L1), composed of custom
hardware, uses information from the calorimeters and muon
detectors to select the most interesting events for more refined selection and analysis at a rate of up to 80~kHz. The software-based High Level Trigger
(HLT) further decreases the rate to a maximum of
$\sim 300$~Hz for data storage. The instantaneous luminosity is measured using information from forward hadronic calorimeters~\cite{lumi}.

We use data collected with a dedicated trigger on the total jet activity, $H_T$, where $H_T$ is defined as the scalar sum of the transverse energies $E_T$ of the jets above a preprogrammed threshold. At L1 this jet $E_T$ threshold was 10~GeV, and the $H_T$ threshold was 50 GeV. At HLT, the jet $E_T$ threshold varied between 20 and 30~GeV, and the $H_T$ threshold between 100 and 200~GeV. The trigger is fully efficient for the offline analysis selections described below. Energetic electrons and photons are also reconstructed as jets at the trigger level and are thus included in the $H_T$ sum.

Jets are reconstructed using energy deposits in the HCAL and ECAL, clustered using a collinear and infrared safe anti-$k_T$ algorithm with a distance parameter of 0.5~\cite{anti-kt}. The jet energy resolution is $\Delta E/E \approx 100\,\%/\sqrt{E\,[\mbox{GeV}]} \oplus 5\,\%$. Jets are required to pass quality requirements to remove those consistent with calorimeter noise. Jet energies are corrected for the non-uniformity and non-linearity of the calorimeter response, as derived using Monte Carlo (MC) samples and collision data~\cite{JES}. Jets are required to have $E_T > 20$~GeV before the jet-energy-scale corrections and to have $|\eta| <  2.6$. Missing transverse energy \MET is reconstructed as the negative of the vector sum of transverse energies in the individual calorimeter towers. This quantity is further corrected to account for muons in the event, which deposit little energy in the calorimeters, and for the jet energy scale~\cite{MET}.

Electrons and photons are identified as isolated energy deposits in the ECAL, with a shape consistent with that expected for electromagnetic showers. Photons are required to have no matching hits in the inner pixel detector layers, while electrons are required to have a matching track. Electrons and photons are required to have $E_T > 20$~GeV and to be reconstructed in the fiducial volume of the barrel ($|\eta| < 1.44$) or the endcap ($1.56 < |\eta| < 2.4$). The ECAL has an ultimate energy resolution better than 0.5\% for unconverted photons or electrons with transverse energies above 100 GeV~\cite{ECAL}. In 2010 collision data, for $E_T > 20$ GeV, this resolution is better than 1\% in the barrel.

Muons are required to have matched tracks in the central tracker and the muon spectrometer, to be within $|\eta| < 2.1$, be consistent with the interaction vertex to suppress backgrounds from cosmic ray muons, be isolated from other tracks, and have transverse momentum $p_T$ above 20~GeV. The combined fit using tracks measured in the central tracker and the muon spectrometer results in $p_T$ resolution between 1\% and 5\% for $p_T$ values up to 1~TeV.

The separation between any two objects (jet, lepton, or photon) is required to be $$\Delta R = \sqrt{\Delta \phi^2 + \Delta \eta^2} > 0.3.$$

Black hole signal events are simulated using the parton-level BlackMax~\cite{BlackMax} generator (v2.01.03), followed by a parton-showering fragmentation with \PYTHIA~\cite{PYTHIA} (v6.420), and a fast parametric simulation of the CMS detector response~\cite{FastSim}, which has been extensively validated for signal events using detailed detector simulation via \GEANTfour~\cite{GEANT4}.

Several additional parameters govern black hole production and decay in the ADD model in addition to $M_D$ and $n$. For each value of $M_D$, we consider a range of the minimum black hole masses, $M_{\rm BH}^{\rm min}$, between $M_D$ and the kinematic limit of the LHC. We assume that no parton-collision energy is lost in gravitational shock waves, i.e. it is all trapped within the event horizon of the forming black hole. We consider both rotating and non-rotating black holes in this analysis. Graviton radiation by the black hole is not considered. For most of the signal samples we assume full Hawking evaporation without a stable non-interacting remnant.

The parameters used in the simulations are listed in Table~\ref{tab:BH} for a number of characteristic model points. The MSTW2008lo68~\cite{MSTW} parton distribution functions (PDF) were used. In addition we compare the BlackMax results with those of the {\sc CHARYBDIS 2} MC generator (v1.0.3)~\cite{CHARYBDIS,CHARYBDIS2}. The two generators yield different values of total cross section, as BlackMax introduces an additional $n$-dependent factor applied on top of the geometrical cross section. The {\sc CHARYBDIS} cross sections are a factor of 1.36, 1.59, and 1.78 smaller than those from BlackMax for $n = 2$, 4, and 6, respectively. In addition, {\sc CHARYBDIS} has been used to simulate black hole evaporation resulting in a stable non-interacting remnant with mass $M_D$ (this model is not implemented in BlackMax). In the generation, we use the Particle Data Group~\cite{PDG} definition of the Planck scale $M_D$. (Using another popular choice for $M_D$ from Dimopoulos and Landsberg~\cite{dl} would result in a suppression of the production cross section by a factor of 1.35, 5.21, or 9.29 for $n = 2$, 4, or 6, respectively.)

\begin{table*}[htbp]
\begin{center}
\caption{Monte Carlo signal points for some of the model parameters probed, corresponding leading order cross sections ($\sigma$), and the minimum required values for the event multiplicity ($N \geq N^{\min}$) and $S_T$ ($S_{T}^{\rm min}$), as well as the signal acceptance ($A$), the expected number of signal events ($n^{\rm sig}$), the number of observed events ($n^{\rm data}$) in data, the expected number of background events ($n^{\rm bkg}$), and the observed ($\sigma^{95}$) and expected ($\sigma^{95}_{\rm exp.}$) limits on the signal cross section at 95\% confidence level.}
\medskip
\label{tab:BH}
\begin{tabular}{c|c|c|c|c|c|c|c|c|c|c|c}
\hline
$M_D$ & $M_{BH}$ & $n$ & $\sigma$ & $N^{\rm min}$ & $S_{T}^{\rm min}$ & A & $n^{\rm sig}$ & $n^{\rm data}$ & $n^{\rm bkg}$ & $\sigma^{95}$ & $\sigma^{95}_{\rm exp.}$ \\
(TeV) & (TeV) & & (pb) & & (TeV) & (\%) & & & & (pb) & (pb)\\
\hline
1.5 & 2.5 & 6 & 117.9 & 3 & 1.5 & 90.6 & 3713 & 203 & 241 $\pm$ 45 & 1.69 & 2.52\\
1.5 & 3.0 & 6 & 25.94 & 3 & 1.8 & 91.3 & 823 & 45 & 66.2 $\pm$ 22.2 & 0.62 & 1.13\\
1.5 & 3.5 & 6 & 4.97 & 4 & 2.1 & 88.3 & 153 & 6 & 12.1 $\pm$ 6.3 & 0.21 & 0.39\\
1.5 & 4.0 & 6 & 0.77 & 5 & 2.4 & 84.4 & 22.5 & 0 & 2.01 $\pm$ 1.48 & 0.11 & 0.18\\
1.5 & 4.5 & 6 & 0.09 & 5 & 2.9 & 80.9 & 2.55 & 0 & 0.46$^{+0.54}_{-0.46}$ & 0.11 & 0.13\\
1.5 & 5.0 & 6 & 0.007 & 5 & 3.4 & 75.2 & 0.19 & 0 & 0.13$^{+0.21}_{-0.13}$ & 0.12 & 0.13\\
2.0 & 2.5 & 4 & 28.88 & 3 & 1.7 & 81.4 & 817 & 82 & 99.7 $\pm$ 28.1 & 1.16 & 1.64\\
2.0 & 3.0 & 4 & 6.45 & 3 & 2.0 & 83.2 & 186 & 21 & 30.8 $\pm$ 14.0 & 0.47 & 0.76\\
2.0 & 3.5 & 4 & 1.26 & 4 & 2.3 & 77.9 & 34.0 & 3 & 6.12 $\pm$ 4.05 & 0.20 & 0.31\\
2.0 & 4.0 & 4 & 0.20 & 4 & 2.8 & 73.4 & 5.07 & 0 & 1.35$^{+1.45}_{-1.35}$ & 0.12 & 0.19\\
2.0 & 4.5 & 4 & 0.02 & 5 & 3.2 & 64.4 & 0.53 & 0 & 0.21$^{+0.31}_{-0.21}$ & 0.14 & 0.15\\
2.0 & 5.0 & 4 & 0.002 & 5 & 3.7 & 59.6 & 0.04 & 0 & 0.06$^{+0.12}_{-0.06}$ & 0.15 & 0.15\\
3.0 & 3.0 & 2 & 0.59 & 3 & 2.4 & 62.1 & 12.8 & 2 & 7.88 $\pm$ 5.80 & 0.21 & 0.46\\
3.0 & 3.5 & 2 & 0.12 & 3 & 2.8 & 58.9 & 2.41 & 0 & 2.40$^{+2.57}_{-2.40}$ & 0.15 & 0.28\\
3.0 & 4.0 & 2 & 0.02 & 4 & 3.2 & 47.3 & 0.32 & 0 & 0.46$^{+0.67}_{-0.46}$ & 0.19 & 0.23\\
3.0 & 4.5 & 2 & 0.002 & 5 & 3.6 & 33.6 & 0.03 & 0 & 0.08$^{+0.15}_{-0.08}$ & 0.26 & 0.28\\
3.0 & 5.0 & 2 & 0.0002 & 5 & 4.0 & 34.5 & 0.002 & 0 & 0.03$^{+0.07}_{-0.03}$ & 0.26 & 0.26\\
\hline
\end{tabular}
\end{center}
\end{table*}

We employ a selection based on total transverse energy to separate black hole candidate events from the backgrounds. The variable $S_T$ is defined as a scalar sum of the $E_T$ of the $N$ individual objects (jets, electrons, photons, and muons) passing the above selections. Only objects with $E_T > 50$~GeV are included in the calculation of $S_T$, in order to suppress the SM backgrounds and to be insensitive to jets from pile-up, while being fully efficient for black hole decays. Further, the missing transverse energy in the event is added to $S_T$, if the missing transverse energy value exceeds 50~GeV. Note that while \MET\ is counted toward $S_T$, it is not considered in the determination of $N$.

The main background to black hole signals arises from QCD multijet events. Other backgrounds from direct photon, $W/Z+$jets, and $t\bar{t}$ production were estimated from MC simulations, using the \MADGRAPH~\cite{MadGraph} leading-order parton-level event generator with CTEQ6L PDF set~\cite{CTEQ}, followed by \PYTHIA~\cite{PYTHIA} parton showering and full CMS detector simulation via \GEANTfour \cite{GEANT4}. These additional backgrounds are negligible at large values of $S_T$ and contribute less than 1\% to the total background after the final selection.

The dominant multijet background can only be estimated reliably from data. For QCD events, $S_T$ is almost completely determined by the hard $2 \to 2$ parton scattering process. Further splitting of the jets due to final-state radiation, as well as jets due to initial-state radiation~-- most often nearly collinear with either incoming or outgoing partons~-- does not change the $S_T$ value considerably. Consequently, the shape of the $S_T$ distribution is expected to be independent of the event multiplicity $N$, as long as $S_T$ is sufficiently above the turn-on region (i.e., much higher than $N \times 50$~GeV).

We confirmed the assumption of the $S_T$ shape invariance of $N$ using MC generators capable of simulating multijet final states from either matrix elements~\cite{ALPGEN} or parton showers~\cite{PYTHIA}. This shape invariance offers a direct way of extracting the expected number of background events in the search for black hole production. The decay of black holes with a mass of a few TeV typically results in events with multiplicity of half-a-dozen objects in the final state. The conjecture that the $S_T$ shape is independent of the multiplicity has also been checked with data using the exclusive multiplicities of $N = 2$ and $N = 3$. Even in the presence of a signal, its contribution to these multiplicities is expected to be small and only seen at large values of $S_T$, so these samples can be used for the background prediction at higher multiplicities. Moreover, since dedicated analyses of the dijet invariant mass spectrum have been conducted~\cite{dijets1,dijets2}, we know that there are no appreciable contributions from new physics to the dijet final state up to invariant masses of about 1.5~TeV, which, for central jets, translates to a similar range of $S_T$.

\begin{figure*}[htbp]
\centering
\includegraphics[width=0.49\textwidth]{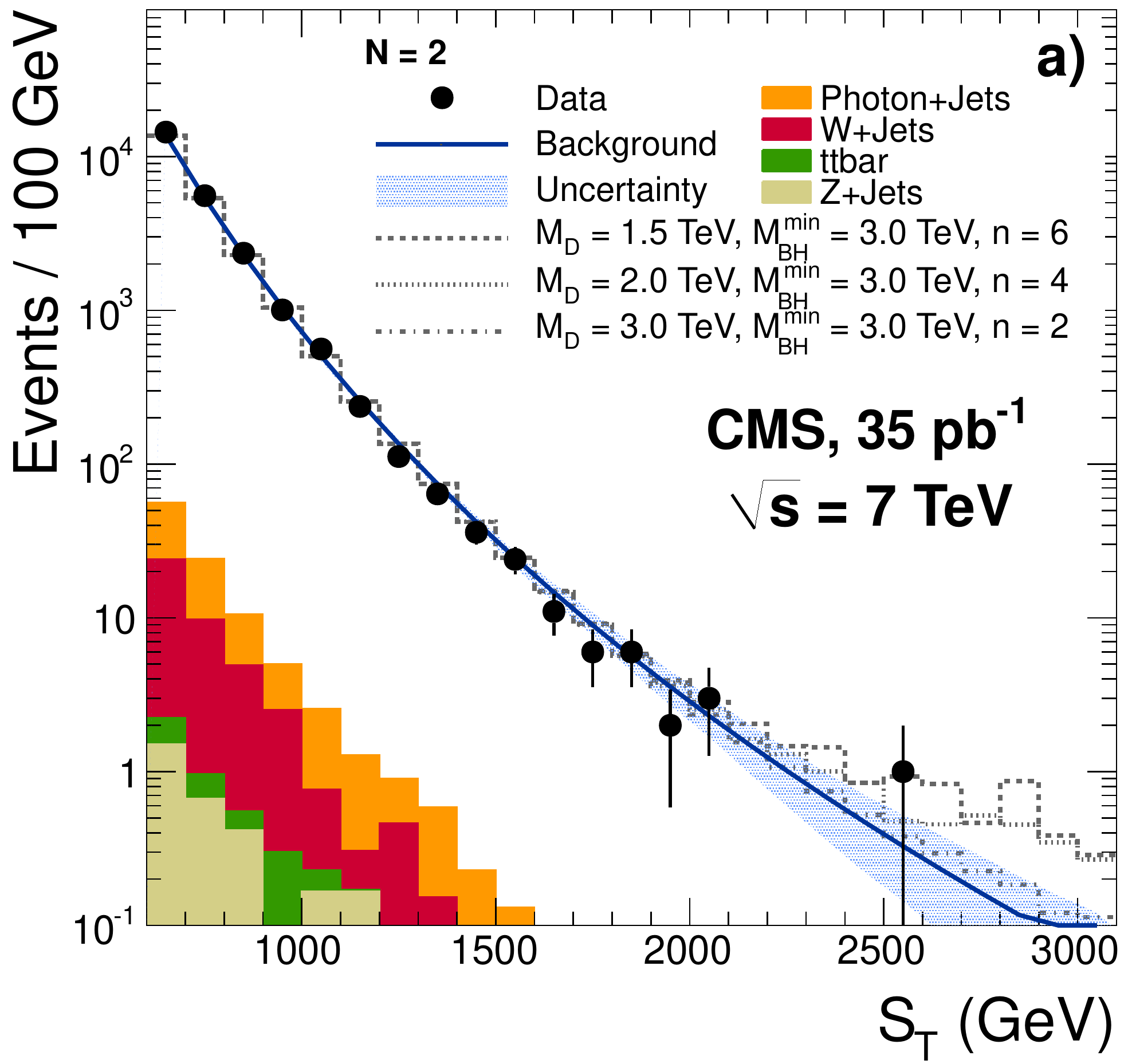}
\includegraphics[width=0.49\textwidth]{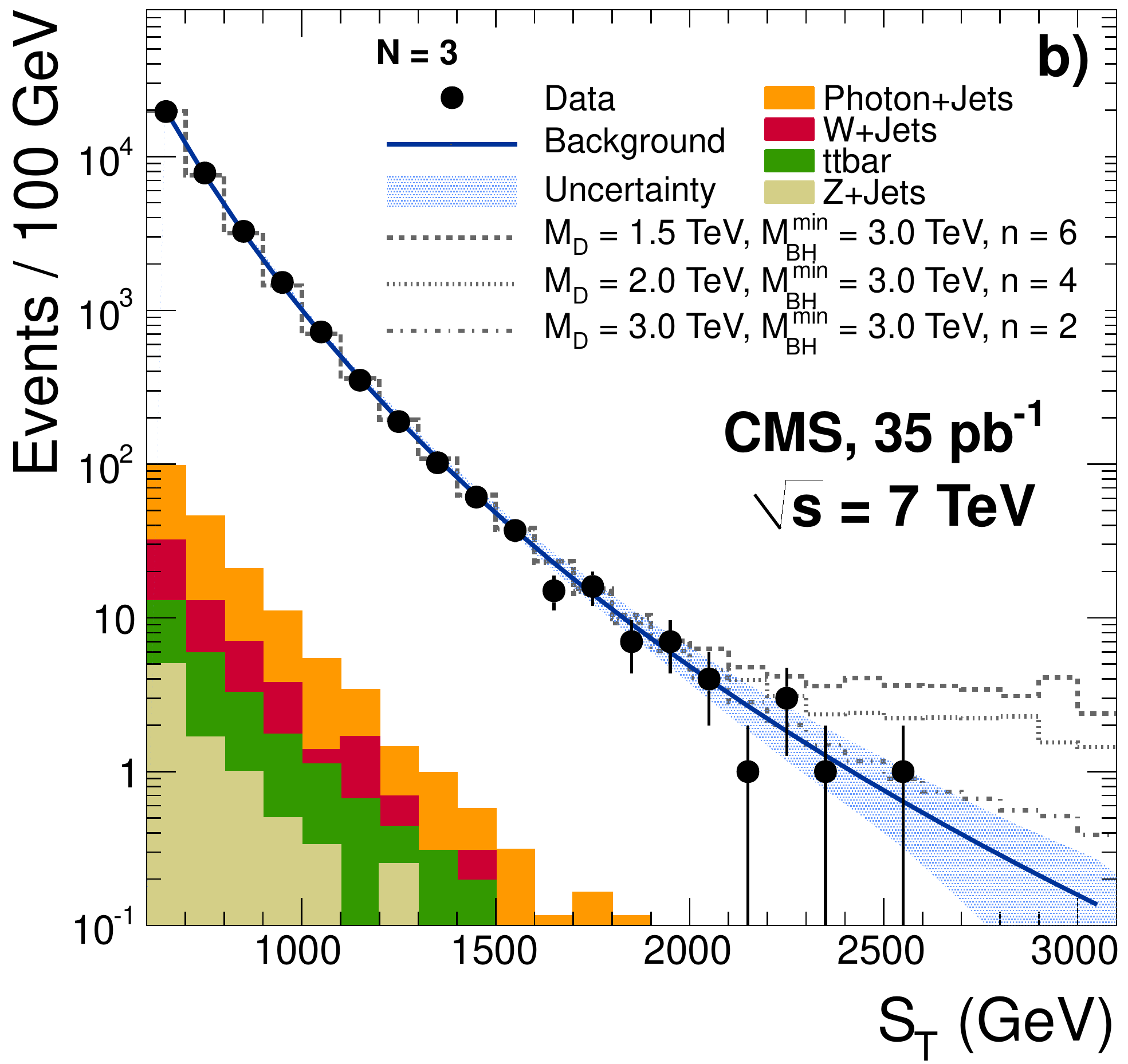}
\caption{Total transverse energy $S_T$, for events with the multiplicities of a) $N = 2$, and b) $N = 3$ objects in the final state. Data are depicted as solid circles with error bars; the shaded band is the
background prediction obtained from data (solid line) with its uncertainty. Non-multijet backgrounds are shown as colored histograms.
Also shown is the predicted black hole signal for three different parameter sets.}
\label{fig:STN}
\end{figure*}

We fit the $S_T$ distributions between 600 and 1100 GeV, where no black hole signal is expected, for data events with $N = 2$ and $N = 3$ using an ansatz function $\frac{P_0 (1 + x)^{P_1}}{x^{P_2 + P_3 \log(x)}}$, which is shown with the solid line in Fig.~\ref{fig:STN}. To check the systematic uncertainty of the fit, we use two additional ansatz functions, $\frac{P_0}{(P_1 + P_2 x + x^2)^{P_3}}$ and $\frac{P_0}{(P_1 + x)^{P_2}}$~\cite{dijets1}, which are shown as the upper and lower boundaries of the shaded band in Fig.~\ref{fig:STN}. The default choice of the ansatz function was made based on the best-fit to the $S_T$ distribution for $N=2$. Additional systematic uncertainty arises from a slight difference between the best-fit shapes for $N = 2$ and $N = 3$. Nevertheless, the fits for these two exclusive multiplicities agree with each other within the uncertainties, demonstrating that the shape of the $S_T$ distribution is independent of the final-state multiplicity.

\begin{figure*}[htbp]
\centering
\includegraphics[width=0.49\textwidth]{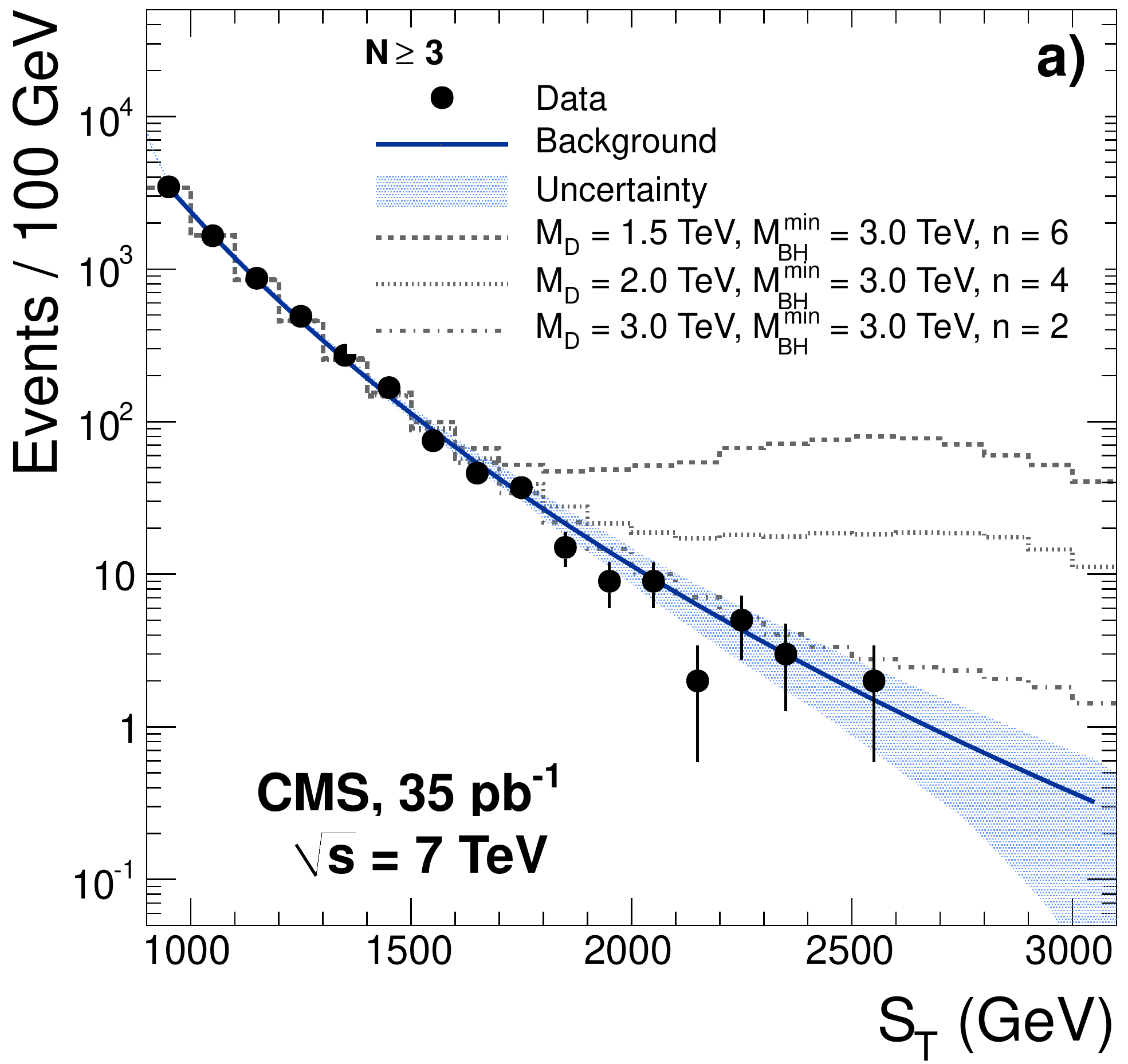}
\includegraphics[width=0.49\textwidth]{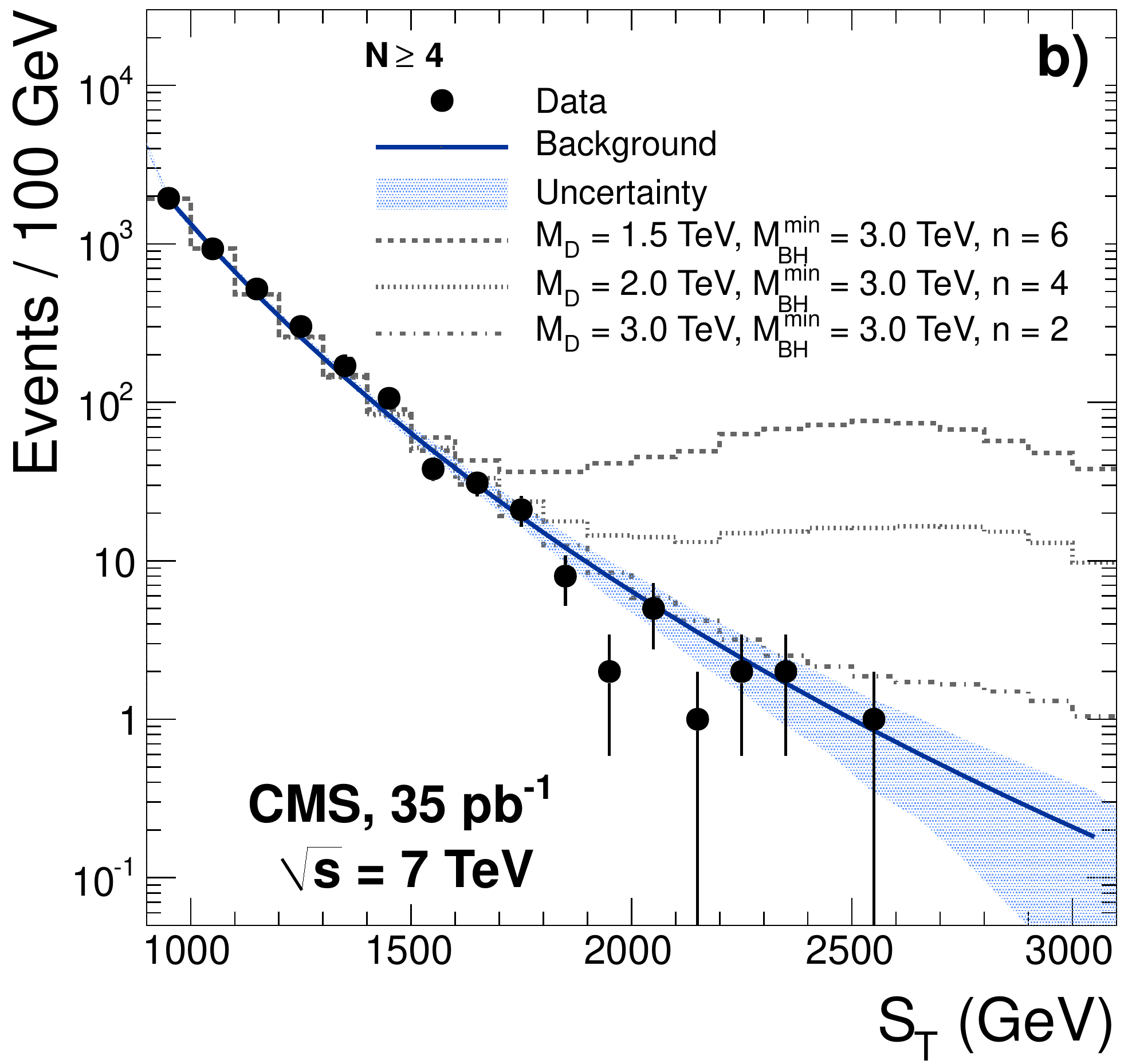}
\includegraphics[width=0.49\textwidth]{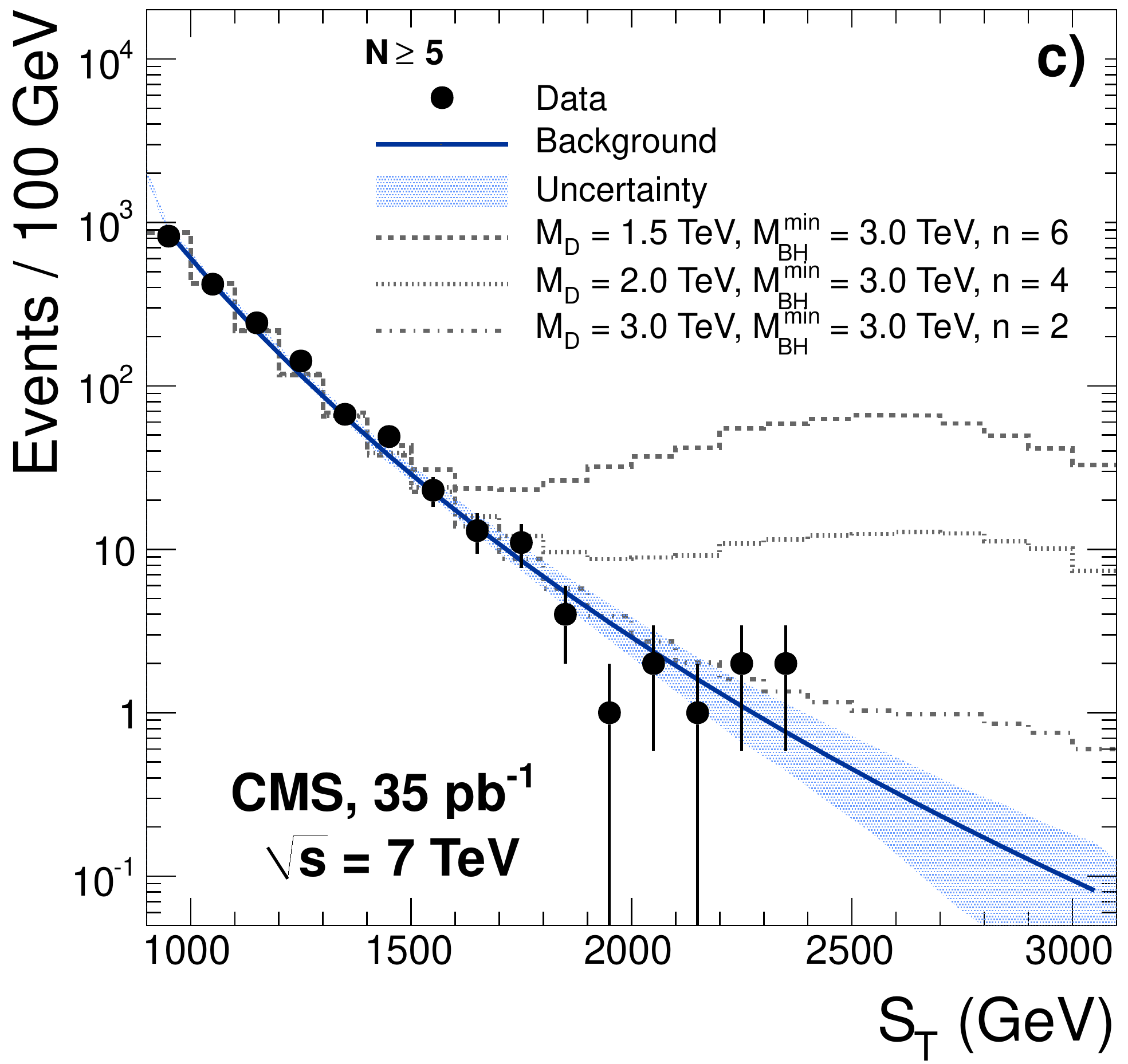}
\caption{Total transverse energy $S_T$, for events with multiplicities a) $N \geq 3$, b) $N \geq 4$, and c) $N \geq 5$ objects in the final state. Data are depicted as solid circles with error bars; the shaded band is the background prediction (solid line) with its uncertainty. Also shown are black hole signals for three different parameter sets.}
\label{fig:STinclusive}
\end{figure*}

The $S_T$ distributions for data events with multiplicities $N \ge 3$, 4, and 5 are shown in Figs.~\ref{fig:STinclusive}a, b, and c, respectively. The solid curves in the figures are the predicted background shapes, found by normalizing the fits of the $N = 2$ and 3 $S_T$ distributions to the range $S_T = 1000 - 1100$~GeV, where no black hole signal contribution is expected.

\begin{figure}[htbp]
\centering
\includegraphics[width=0.6\textwidth]{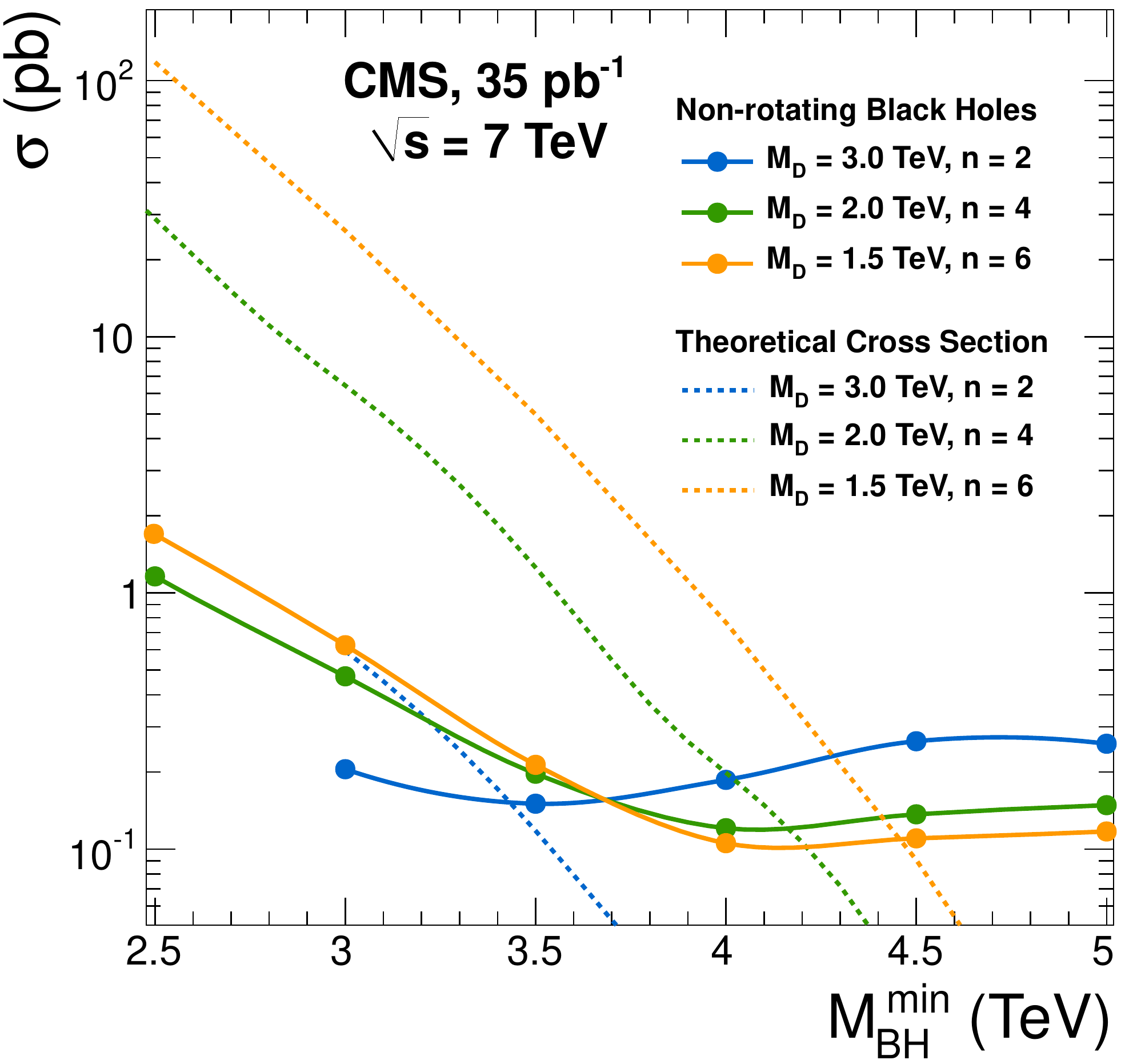}
\caption{The 95\% confidence level upper limits on the black hole production cross section (solid lines) and three theoretical predictions for the cross section (dotted lines), as a function of the black hole mass.}
\label{fig:BHL}
\end{figure}

\begin{figure}[htbp]
\centering
\includegraphics[width=0.6\textwidth]{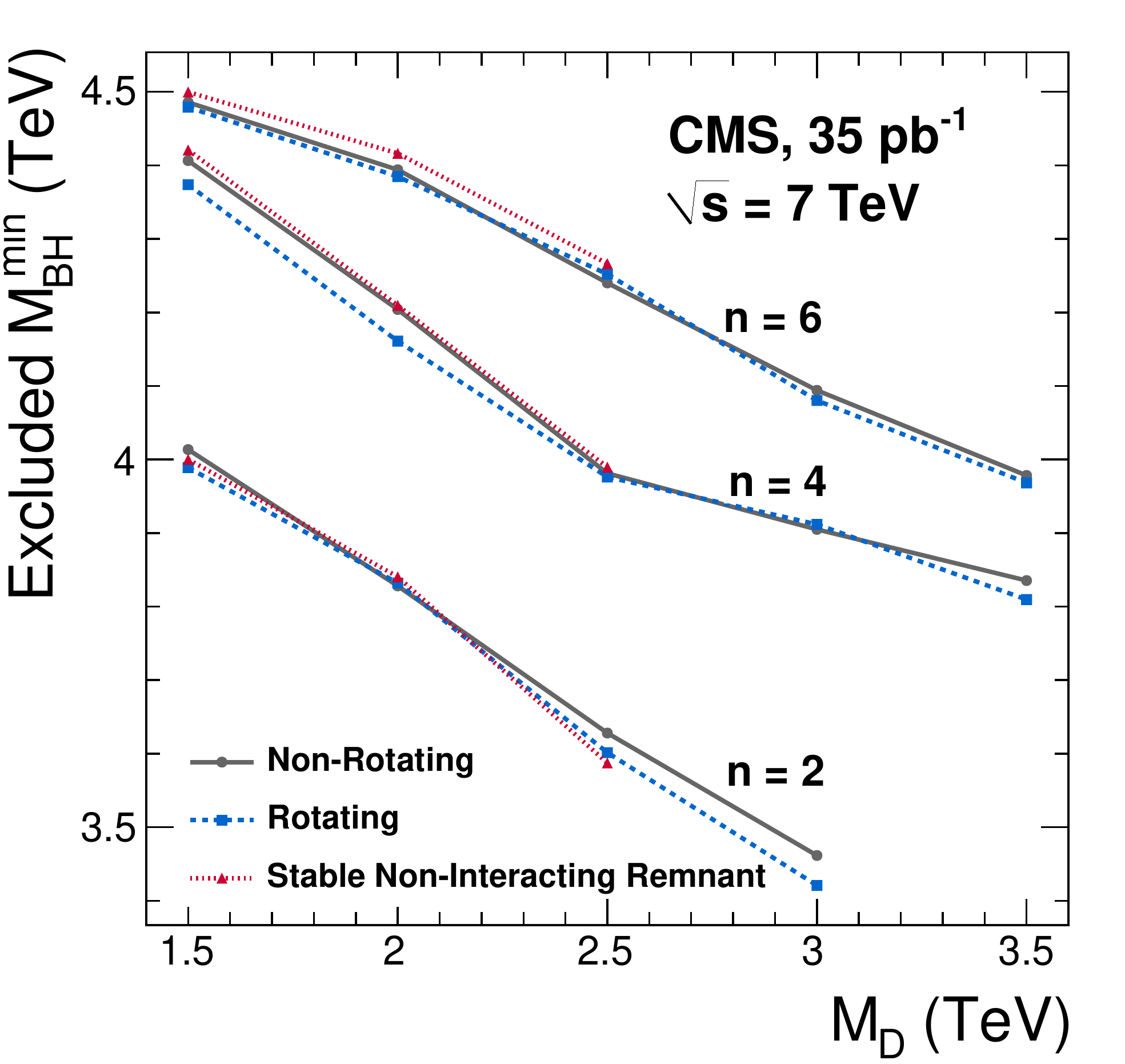}
\caption{The 95\% confidence level limits on the black hole mass as a function of the multidimensional Planck scale $M_D$ for several benchmark scenarios. The area below each curve is excluded by this search.}
\label{fig:BHpar}
\end{figure}

Since no excess is observed above the predicted background, we set limits on the black hole production. We assign a systematic uncertainty on the background estimate of 6\% to 125\% for the $S_T$ range used in this search. This uncertainy comes from the normalization uncertainty (4 -- 12\%, dominated by the statistics in the normalization region) added in quadrature to the uncertainties arising from using various ansatz fit functions and the difference between the shapes obtained from the $N = 2$ and $N = 3$ samples. The integrated luminosity is measured with an uncertainty of 11\%~\cite{lumi}. The uncertainty on the signal yield is dominated by the jet energy scale uncertainty of $\approx 5\%$~\cite{JES} which translates into a 5\% uncertainty on the signal. An additional 2\% uncertainty on the signal acceptance comes from the variation of PDFs within the CTEQ6 error set~\cite{CTEQ}. The particle identification efficiency does not affect the signal distribution, since an electron failing the identification requirements would be classified either as a photon or a jet; a photon failing the selection would become a jet; a rejected muon would contribute to the \MET. In any case the total value of $S_T$ is not affected.

We set limits on black hole production with the optimized $S_T$ and $N$ selections by counting events with $S_T > S_{T}^{\rm min}$ and $N > N^{\rm min}$. We optimized the signal ($S$) significance in the presence of background ($B$) using the ratio $S/\sqrt{S+B}$ for each set. The optimum choice of parameters is listed in Table~\ref{tab:BH}, as well as the predicted number of background events, the expected number of signal events, and the observed number of events in data. Note that the background uncertainty, dominated by the choice of the fitting function, is highly correlated for various working points listed in Table~\ref{tab:BH} and also bin-to-bin for the $S_T$ distributions shown in Figs.~\ref{fig:STN} and \ref{fig:STinclusive}.

We set upper limits on the black hole production cross section using the Bayesian method with flat signal prior and log-normal prior for integration over the nuisance parameters (background, signal acceptance, luminosity)~\cite{Bayes,PDG}. These upper limits at the 95\% confidence level (CL) are shown in Fig.~\ref{fig:BHL}, as a function of $M_{\rm BH}^{\rm min}$. For the three working points shown in the figure, the observed (expected) lower limits on the black hole mass are 3.5, 4.2 and 4.5 TeV (3.2, 4.0, and 4.5 TeV), respectively.

Translating these upper limits into lower limits on the parameters of the ADD model, we can exclude the production of black holes with minimum mass of $3.5 - 4.5$~TeV for values of the multidimensional Planck scale up to 3.5~TeV at 95\% CL. These limits, shown in Fig.~\ref{fig:BHpar}, do not exhibit significant dependence on the details of the production and evaporation model. These are the first limits of a dedicated search for black hole production at hadron colliders.

\begin{figure*}[htbp]
\centering
\includegraphics[width=0.32\textwidth]{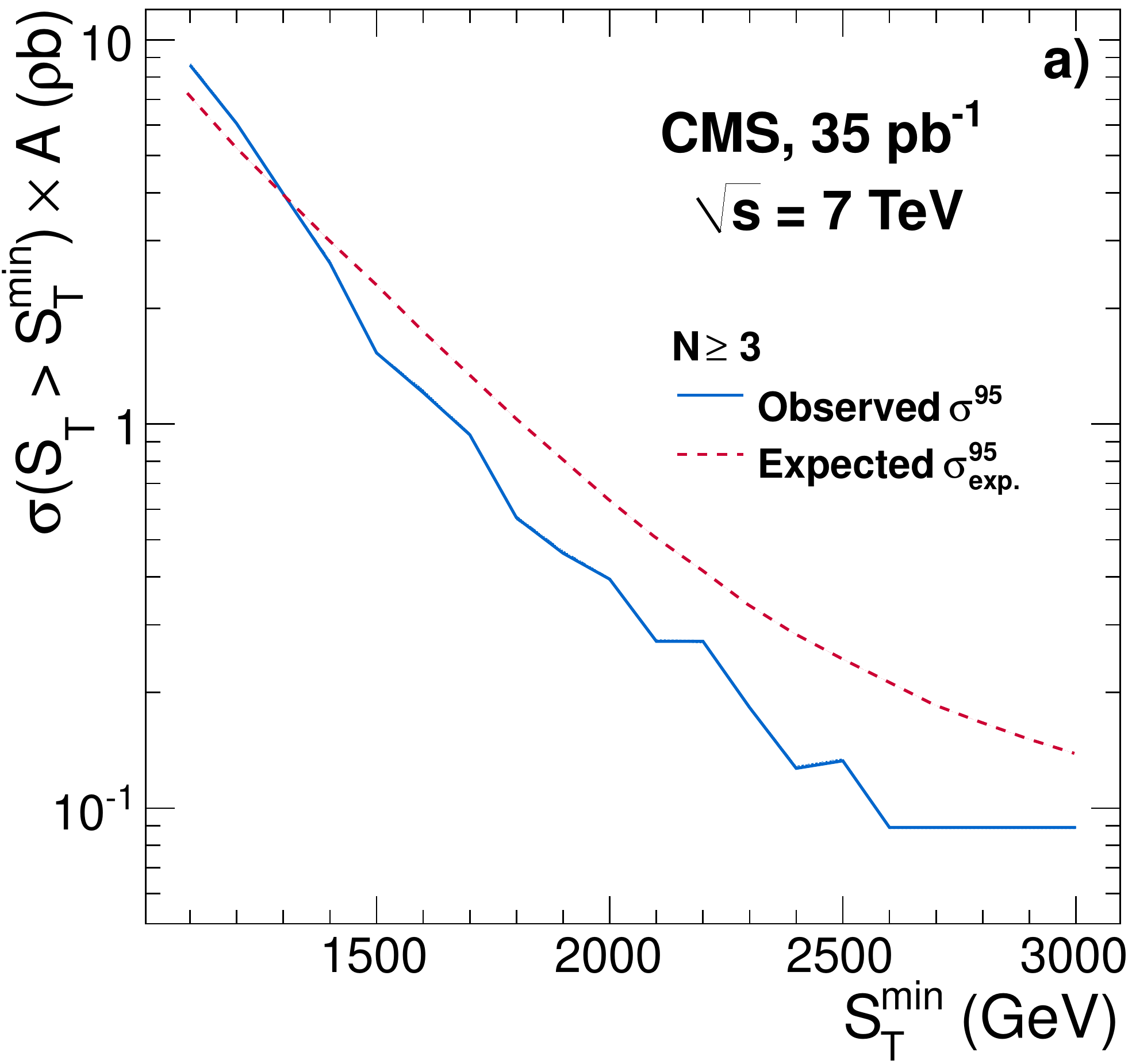}
\includegraphics[width=0.32\textwidth]{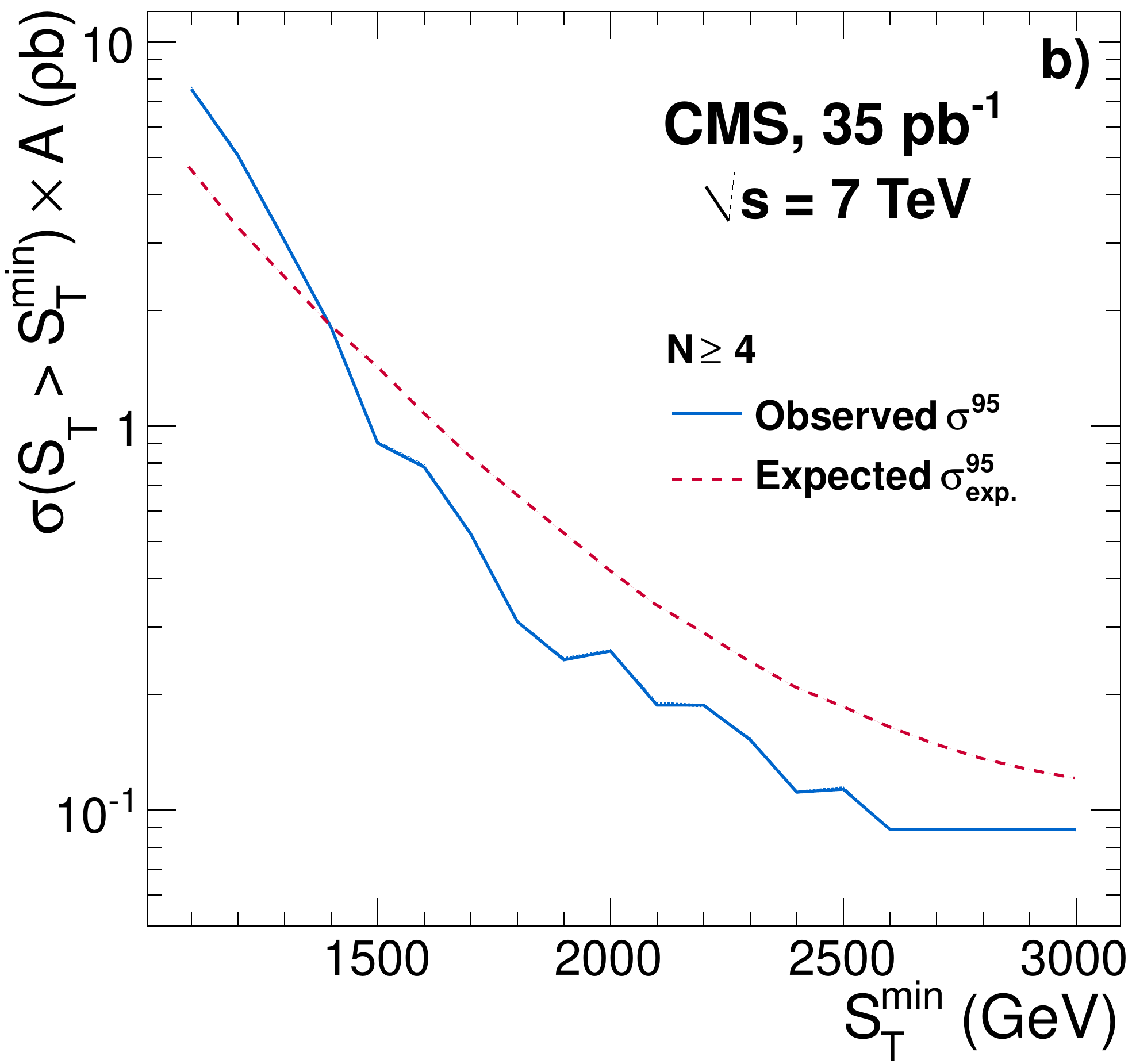}
\includegraphics[width=0.32\textwidth]{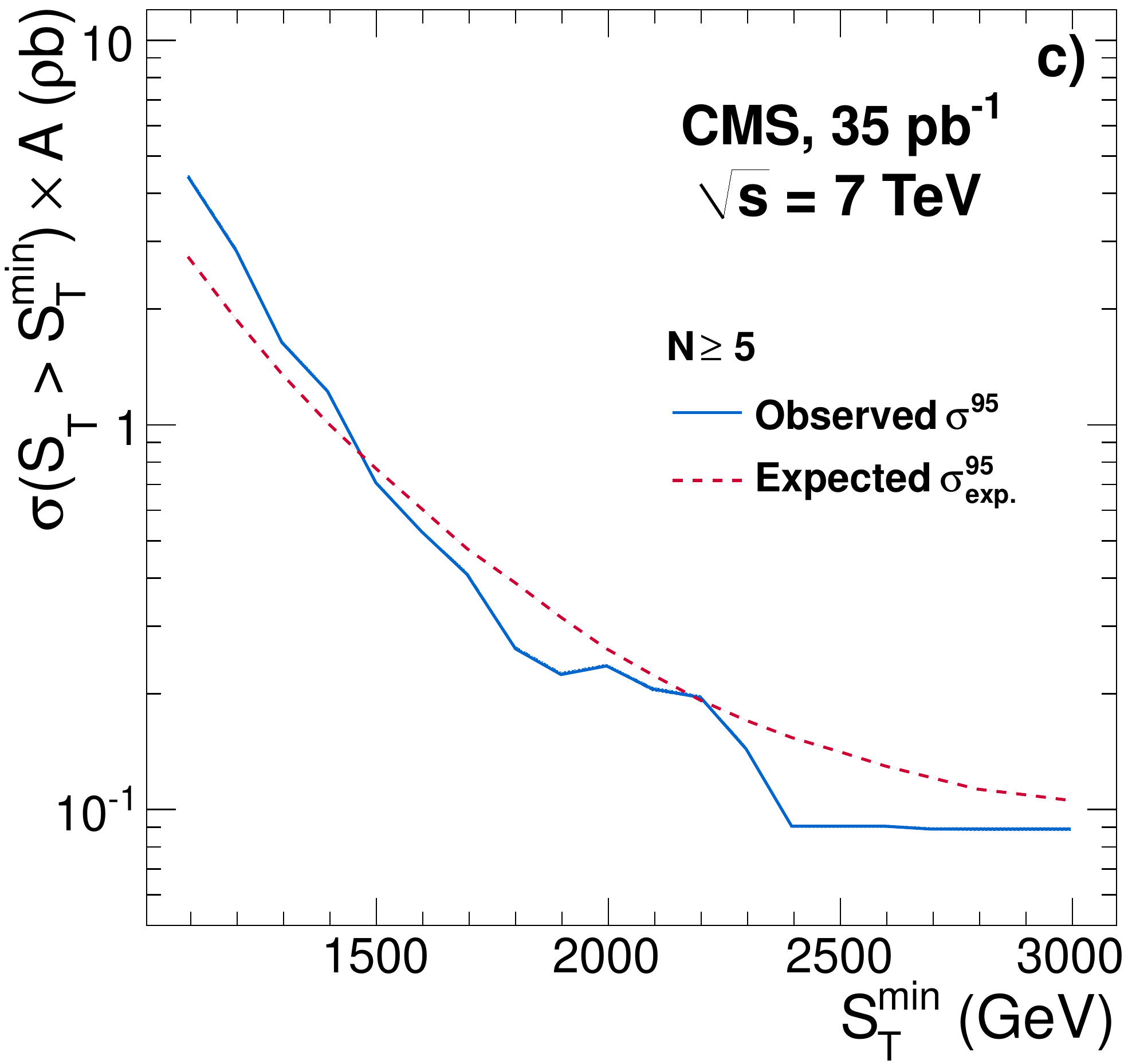}
\caption{Model-independent 95\% confidence level upper limits on a signal cross section times acceptance for counting experiments with $S_T > S_T^{\rm min}$ as a function of $S_T^{\rm min}$ for (a) $N \ge 3$, (b) $N \ge 4$, and (c) $N \ge 5$. The blue (red) lines correspond to an observed (expected) limit for nominal signal acceptance uncertainty of 5\%.}
\label{fig:modind_limits}
\end{figure*}

Finally, we produce model-independent upper limits on the cross section times the acceptance for new physics production in high-$S_T$ inclusive final states for $N \geq 3$, 4, and 5. Figure~\ref{fig:modind_limits} shows 95\% CL upper limits from a counting experiment for $S_T > S_T^{\rm min}$ as a function of $S_T^{\rm min}$, which can be used to test models of new physics that result in these final states. A few examples of such models are production of high-mass $t\bar t$ resonances~\cite{ttbar} in the six-jet and lepton + jet final states, $R$-parity violating gluino decay into three jets, resulting in the six-jet final state~\cite{sixjet1,sixjet2}, and a class of models with strong dynamics, with a strongly produced resonance decaying into a pair of resonances further decaying into two jets each, resulting in the four-jet final state~\cite{coloron}. In addition, these limits can be used to constrain black hole production for additional regions of the parameter space of the model, as well as set limits on the existence of string balls~\cite{SB}, which are quantum precursors of black holes predicted in certain string models. We have checked that for the black hole model parameters we probed with the dedicated optimized analysis, the sensitivity of the search in terms of the excluded black hole mass range exceeds that from the model-independent cross section limits by as little as 5 -- 8\%. Thus, model-independent limits can be used efficiently to constrain the allowed parameter space in an even broader variety of black hole models than we covered in this Letter.

To conclude, we have performed the first dedicated search for microscopic black holes at a particle accelerator and set limits on their production within a variety of models. The lower limits on the black hole mass at 95\% CL range from 3.5 to 4.5~TeV for values of the Planck scale up to 3.5~TeV in the model with large extra dimensions in space. Additionally, we have produced model-independent limits on the production of energetic, high-multiplicity final states, which can be used to constrain a variety of models of new physics.

We wish to congratulate our colleagues in the CERN accelerator departments for the excellent performance of the LHC machine. We thank the technical and administrative staff at CERN and other CMS institutes, and acknowledge support from: FMSR (Austria); FNRS and FWO (Belgium); CNPq, CAPES, FAPERJ, and FAPESP (Brazil); MES (Bulgaria); CERN; CAS, MoST, and NSFC (China); COLCIENCIAS (Colombia); MSES (Croatia); RPF (Cyprus); Academy of Sciences and NICPB (Estonia); Academy of Finland, ME, and HIP (Finland); CEA and CNRS/IN2P3 (France); BMBF, DFG, and HGF (Germany); GSRT (Greece); OTKA and NKTH (Hungary); DAE and DST (India); IPM (Iran); SFI (Ireland); INFN (Italy); NRF and WCU (Korea); LAS (Lithuania); CINVESTAV, CONACYT, SEP, and UASLP-FAI (Mexico); PAEC (Pakistan); SCSR (Poland); FCT (Portugal); JINR (Armenia, Belarus, Georgia, Ukraine, Uzbekistan); MST and MAE (Russia); MSTD (Serbia); MICINN and CPAN (Spain); Swiss Funding Agencies (Switzerland); NSC (Taipei); TUBITAK and TAEK (Turkey); STFC (United Kingdom); DOE and NSF (USA).
\bibliography{auto_generated}   
\cleardoublepage\appendix\section{The CMS Collaboration \label{app:collab}}\begin{sloppypar}\hyphenpenalty=5000\widowpenalty=500\clubpenalty=5000\input{EXO-10-017-authorlist.tex}\end{sloppypar}
\end{document}

%% file: ptdr-definitions.tex
%
%
%

\providecommand {\etal}{\mbox{et al.}\xspace} 
\providecommand {\ie}{\mbox{i.e.}\xspace}     
\providecommand {\eg}{\mbox{e.g.}\xspace}     
\providecommand {\etc}{\mbox{etc.}\xspace}     
\providecommand {\vs}{\mbox{\sl vs.}\xspace}      
\providecommand {\mdash}{\ensuremath{\mathrm{-}}} 

\providecommand {\Lone}{Level-1\xspace} 
\providecommand {\Ltwo}{Level-2\xspace}
\providecommand {\Lthree}{Level-3\xspace}

\providecommand{\ACERMC} {\textsc{AcerMC}\xspace}
\providecommand{\ALPGEN} {{\textsc{alpgen}}\xspace}
\providecommand{\CHARYBDIS} {{\textsc{charybdis}}\xspace}
\providecommand{\CMKIN} {\textsc{cmkin}\xspace}
\providecommand{\CMSIM} {{\textsc{cmsim}}\xspace}
\providecommand{\CMSSW} {{\textsc{cmssw}}\xspace}
\providecommand{\COBRA} {{\textsc{cobra}}\xspace}
\providecommand{\COCOA} {{\textsc{cocoa}}\xspace}
\providecommand{\COMPHEP} {\textsc{CompHEP}\xspace}
\providecommand{\EVTGEN} {{\textsc{evtgen}}\xspace}
\providecommand{\FAMOS} {{\textsc{famos}}\xspace}
\providecommand{\GARCON} {\textsc{garcon}\xspace}
\providecommand{\GARFIELD} {{\textsc{garfield}}\xspace}
\providecommand{\GEANE} {{\textsc{geane}}\xspace}
\providecommand{\GEANTfour} {{\textsc{geant4}}\xspace}
\providecommand{\GEANTthree} {{\textsc{geant3}}\xspace}
\providecommand{\GEANT} {{\textsc{geant}}\xspace}
\providecommand{\HDECAY} {\textsc{hdecay}\xspace}
\providecommand{\HERWIG} {{\textsc{herwig}}\xspace}
\providecommand{\HIGLU} {{\textsc{higlu}}\xspace}
\providecommand{\HIJING} {{\textsc{hijing}}\xspace}
\providecommand{\IGUANA} {\textsc{iguana}\xspace}
\providecommand{\ISAJET} {{\textsc{isajet}}\xspace}
\providecommand{\ISAPYTHIA} {{\textsc{isapythia}}\xspace}
\providecommand{\ISASUGRA} {{\textsc{isasugra}}\xspace}
\providecommand{\ISASUSY} {{\textsc{isasusy}}\xspace}
\providecommand{\ISAWIG} {{\textsc{isawig}}\xspace}
\providecommand{\MADGRAPH} {\textsc{MadGraph}\xspace}
\providecommand{\MCATNLO} {\textsc{mc@nlo}\xspace}
\providecommand{\MCFM} {\textsc{mcfm}\xspace}
\providecommand{\MILLEPEDE} {{\textsc{millepede}}\xspace}
\providecommand{\ORCA} {{\textsc{orca}}\xspace}
\providecommand{\OSCAR} {{\textsc{oscar}}\xspace}
\providecommand{\PHOTOS} {\textsc{photos}\xspace}
\providecommand{\PROSPINO} {\textsc{prospino}\xspace}
\providecommand{\PYTHIA} {{\textsc{pythia}}\xspace}
\providecommand{\SHERPA} {{\textsc{sherpa}}\xspace}
\providecommand{\TAUOLA} {\textsc{tauola}\xspace}
\providecommand{\TOPREX} {\textsc{TopReX}\xspace}
\providecommand{\XDAQ} {{\textsc{xdaq}}\xspace}

\providecommand {\DZERO}{D\O\xspace}     


\providecommand{\de}{\ensuremath{^\circ}}
\providecommand{\ten}[1]{\ensuremath{\times \text{10}^\text{#1}}}
\providecommand{\unit}[1]{\ensuremath{\text{\,#1}}\xspace}
\providecommand{\mum}{\ensuremath{\,\mu\text{m}}\xspace}
\providecommand{\micron}{\ensuremath{\,\mu\text{m}}\xspace}
\providecommand{\cm}{\ensuremath{\,\text{cm}}\xspace}
\providecommand{\mm}{\ensuremath{\,\text{mm}}\xspace}
\providecommand{\mus}{\ensuremath{\,\mu\text{s}}\xspace}
\providecommand{\keV}{\ensuremath{\,\text{ke\hspace{-.08em}V}}\xspace}
\providecommand{\MeV}{\ensuremath{\,\text{Me\hspace{-.08em}V}}\xspace}
\providecommand{\GeV}{\ensuremath{\,\text{Ge\hspace{-.08em}V}}\xspace}
\providecommand{\gev}{\GeV}
\providecommand{\TeV}{\ensuremath{\,\text{Te\hspace{-.08em}V}}\xspace}
\providecommand{\PeV}{\ensuremath{\,\text{Pe\hspace{-.08em}V}}\xspace}
\providecommand{\keVc}{\ensuremath{{\,\text{ke\hspace{-.08em}V\hspace{-0.16em}/\hspace{-0.08em}}c}}\xspace}
\providecommand{\MeVc}{\ensuremath{{\,\text{Me\hspace{-.08em}V\hspace{-0.16em}/\hspace{-0.08em}}c}}\xspace}
\providecommand{\GeVc}{\ensuremath{{\,\text{Ge\hspace{-.08em}V\hspace{-0.16em}/\hspace{-0.08em}}c}}\xspace}
\providecommand{\TeVc}{\ensuremath{{\,\text{Te\hspace{-.08em}V\hspace{-0.16em}/\hspace{-0.08em}}c}}\xspace}
\providecommand{\keVcc}{\ensuremath{{\,\text{ke\hspace{-.08em}V\hspace{-0.16em}/\hspace{-0.08em}}c^\text{2}}}\xspace}
\providecommand{\MeVcc}{\ensuremath{{\,\text{Me\hspace{-.08em}V\hspace{-0.16em}/\hspace{-0.08em}}c^\text{2}}}\xspace}
\providecommand{\GeVcc}{\ensuremath{{\,\text{Ge\hspace{-.08em}V\hspace{-0.16em}/\hspace{-0.08em}}c^\text{2}}}\xspace}
\providecommand{\TeVcc}{\ensuremath{{\,\text{Te\hspace{-.08em}V\hspace{-0.16em}/\hspace{-0.08em}}c^\text{2}}}\xspace}

\providecommand{\pbinv} {\mbox{\ensuremath{\,\text{pb}^\text{$-$1}}}\xspace}
\providecommand{\fbinv} {\mbox{\ensuremath{\,\text{fb}^\text{$-$1}}}\xspace}
\providecommand{\nbinv} {\mbox{\ensuremath{\,\text{nb}^\text{$-$1}}}\xspace}
\providecommand{\percms}{\ensuremath{\,\text{cm}^\text{$-$2}\,\text{s}^\text{$-$1}}\xspace}
\providecommand{\lumi}{\ensuremath{\mathcal{L}}\xspace}
\providecommand{\Lumi}{\ensuremath{\mathcal{L}}\xspace}
%
\providecommand{\LvLow}  {\ensuremath{\mathcal{L}=\text{10}^\text{32}\,\text{cm}^\text{$-$2}\,\text{s}^\text{$-$1}}\xspace}
\providecommand{\LLow}   {\ensuremath{\mathcal{L}=\text{10}^\text{33}\,\text{cm}^\text{$-$2}\,\text{s}^\text{$-$1}}\xspace}
\providecommand{\lowlumi}{\ensuremath{\mathcal{L}=\text{2}\times \text{10}^\text{33}\,\text{cm}^\text{$-$2}\,\text{s}^\text{$-$1}}\xspace}
\providecommand{\LMed}   {\ensuremath{\mathcal{L}=\text{2}\times \text{10}^\text{33}\,\text{cm}^\text{$-$2}\,\text{s}^\text{$-$1}}\xspace}
\providecommand{\LHigh}  {\ensuremath{\mathcal{L}=\text{10}^\text{34}\,\text{cm}^\text{$-$2}\,\text{s}^\text{$-$1}}\xspace}
\providecommand{\hilumi} {\ensuremath{\mathcal{L}=\text{10}^\text{34}\,\text{cm}^\text{$-$2}\,\text{s}^\text{$-$1}}\xspace}


\providecommand{\PT}{\ensuremath{p_{\mathrm{T}}}\xspace}
\providecommand{\pt}{\ensuremath{p_{\mathrm{T}}}\xspace}
\providecommand{\ET}{\ensuremath{E_{\mathrm{T}}}\xspace}
\providecommand{\HT}{\ensuremath{H_{\mathrm{T}}}\xspace}
\providecommand{\et}{\ensuremath{E_{\mathrm{T}}}\xspace}
\providecommand{\Em}{\ensuremath{E\hspace{-0.6em}/}\xspace}
\providecommand{\Pm}{\ensuremath{p\hspace{-0.5em}/}\xspace}
\providecommand{\PTm}{\ensuremath{{p}_\mathrm{T}\hspace{-1.02em}/}\xspace}
\providecommand{\PTslash}{\ensuremath{{p}_\mathrm{T}\hspace{-1.02em}/}\xspace}
\providecommand{\ETm}{\ensuremath{E_{\mathrm{T}}^{\text{miss}}}\xspace}
\providecommand{\ETslash}{\ensuremath{E_{\mathrm{T}}\hspace{-1.1em}/}\xspace}
\providecommand{\MET}{\ensuremath{E_{\mathrm{T}}^{\text{miss}}}\xspace}
\providecommand{\ETmiss}{\ensuremath{E_{\mathrm{T}}^{\text{miss}}}\xspace}
\providecommand{\VEtmiss}{\ensuremath{{\vec E}_{\mathrm{T}}^{\text{miss}}}\xspace}

\providecommand{\dd}[2]{\ensuremath{\frac{\mathrm{d} #1}{\mathrm{d} #2}}}

\ifthenelse{\boolean{cms@italic}}{\newcommand{\cmsSymbolFace}{\relax}}{\newcommand{\cmsSymbolFace}{\mathrm}}

\providecommand{\zp}{\ensuremath{\cmsSymbolFace{Z}^\prime}\xspace}
\providecommand{\JPsi}{\ensuremath{\cmsSymbolFace{J}\hspace{-.08em}/\hspace{-.14em}\psi}\xspace}


\providecommand{\AFB}{\ensuremath{A_\text{FB}}\xspace}
\providecommand{\wangle}{\ensuremath{\sin^{2}\theta_{\text{eff}}^\text{lept}(M^2_\mathrm{Z})}\xspace}
\providecommand{\stat}{\ensuremath{\,\text{(stat.)}}\xspace}
\providecommand{\syst}{\ensuremath{\,\text{(syst.)}}\xspace}
\providecommand{\kt}{\ensuremath{k_{\mathrm{T}}}\xspace}

\providecommand{\BC}{\ensuremath{\mathrm{B_{c}}}\xspace}
\providecommand{\bbarc}{\ensuremath{\mathrm{\overline{b}c}}\xspace}
\providecommand{\bbbar}{\ensuremath{\mathrm{b\overline{b}}}\xspace}
\providecommand{\ccbar}{\ensuremath{\mathrm{c\overline{c}}}\xspace}
\providecommand{\bspsiphi}{\ensuremath{\mathrm{B_s} \to \JPsi\, \phi}\xspace}
\providecommand{\EE}{\ensuremath{\mathrm{e^+e^-}}\xspace}
\providecommand{\MM}{\ensuremath{\mu^+\mu^-}\xspace}
\providecommand{\TT}{\ensuremath{\tau^+\tau^-}\xspace}
\providecommand{\ttbar}{\ensuremath{\mathrm{t\overline{t}}}\xspace}

\providecommand{\HGG}{\ensuremath{\mathrm{H}\to\gamma\gamma}}
\providecommand{\GAMJET}{\ensuremath{\gamma + \text{jet}}}
\providecommand{\PPTOJETS}{\ensuremath{\mathrm{pp}\to\text{jets}}}
\providecommand{\PPTOGG}{\ensuremath{\mathrm{pp}\to\gamma\gamma}}
\providecommand{\PPTOGAMJET}{\ensuremath{\mathrm{pp}\to\gamma + \mathrm{jet}}}
\providecommand{\MH}{\ensuremath{M_{\mathrm{H}}}}
\providecommand{\RNINE}{\ensuremath{R_\mathrm{9}}}
\providecommand{\DR}{\ensuremath{\Delta R}}

%

\providecommand{\ga}{\ensuremath{\gtrsim}}
\providecommand{\la}{\ensuremath{\lesssim}}
\providecommand{\swsq}{\ensuremath{\sin^2\theta_\mathrm{W}}\xspace}
\providecommand{\cwsq}{\ensuremath{\cos^2\theta_\mathrm{W}}\xspace}
\providecommand{\tanb}{\ensuremath{\tan\beta}\xspace}
\providecommand{\tanbsq}{\ensuremath{\tan^{2}\beta}\xspace}
\providecommand{\sidb}{\ensuremath{\sin 2\beta}\xspace}
\providecommand{\alpS}{\ensuremath{\alpha_S}\xspace}
\providecommand{\alpt}{\ensuremath{\tilde{\alpha}}\xspace}

\providecommand{\QL}{\ensuremath{\mathrm{Q}_\mathrm{L}}\xspace}
\providecommand{\sQ}{\ensuremath{\tilde{\mathrm{Q}}}\xspace}
\providecommand{\sQL}{\ensuremath{\tilde{\mathrm{Q}}_\mathrm{L}}\xspace}
\providecommand{\ULC}{\ensuremath{\mathrm{U}_\mathrm{L}^\mathrm{C}}\xspace}
\providecommand{\sUC}{\ensuremath{\tilde{\mathrm{U}}^\mathrm{C}}\xspace}
\providecommand{\sULC}{\ensuremath{\tilde{\mathrm{U}}_\mathrm{L}^\mathrm{C}}\xspace}
\providecommand{\DLC}{\ensuremath{\mathrm{D}_\mathrm{L}^\mathrm{C}}\xspace}
\providecommand{\sDC}{\ensuremath{\tilde{\mathrm{D}}^\mathrm{C}}\xspace}
\providecommand{\sDLC}{\ensuremath{\tilde{\mathrm{D}}_\mathrm{L}^\mathrm{C}}\xspace}
\providecommand{\LL}{\ensuremath{\mathrm{L}_\mathrm{L}}\xspace}
\providecommand{\sL}{\ensuremath{\tilde{\mathrm{L}}}\xspace}
\providecommand{\sLL}{\ensuremath{\tilde{\mathrm{L}}_\mathrm{L}}\xspace}
\providecommand{\ELC}{\ensuremath{\mathrm{E}_\mathrm{L}^\mathrm{C}}\xspace}
\providecommand{\sEC}{\ensuremath{\tilde{\mathrm{E}}^\mathrm{C}}\xspace}
\providecommand{\sELC}{\ensuremath{\tilde{\mathrm{E}}_\mathrm{L}^\mathrm{C}}\xspace}
\providecommand{\sEL}{\ensuremath{\tilde{\mathrm{E}}_\mathrm{L}}\xspace}
\providecommand{\sER}{\ensuremath{\tilde{\mathrm{E}}_\mathrm{R}}\xspace}
\providecommand{\sFer}{\ensuremath{\tilde{\mathrm{f}}}\xspace}
\providecommand{\sQua}{\ensuremath{\tilde{\mathrm{q}}}\xspace}
\providecommand{\sUp}{\ensuremath{\tilde{\mathrm{u}}}\xspace}
\providecommand{\suL}{\ensuremath{\tilde{\mathrm{u}}_\mathrm{L}}\xspace}
\providecommand{\suR}{\ensuremath{\tilde{\mathrm{u}}_\mathrm{R}}\xspace}
\providecommand{\sDw}{\ensuremath{\tilde{\mathrm{d}}}\xspace}
\providecommand{\sdL}{\ensuremath{\tilde{\mathrm{d}}_\mathrm{L}}\xspace}
\providecommand{\sdR}{\ensuremath{\tilde{\mathrm{d}}_\mathrm{R}}\xspace}
\providecommand{\sTop}{\ensuremath{\tilde{\mathrm{t}}}\xspace}
\providecommand{\stL}{\ensuremath{\tilde{\mathrm{t}}_\mathrm{L}}\xspace}
\providecommand{\stR}{\ensuremath{\tilde{\mathrm{t}}_\mathrm{R}}\xspace}
\providecommand{\stone}{\ensuremath{\tilde{\mathrm{t}}_1}\xspace}
\providecommand{\sttwo}{\ensuremath{\tilde{\mathrm{t}}_2}\xspace}
\providecommand{\sBot}{\ensuremath{\tilde{\mathrm{b}}}\xspace}
\providecommand{\sbL}{\ensuremath{\tilde{\mathrm{b}}_\mathrm{L}}\xspace}
\providecommand{\sbR}{\ensuremath{\tilde{\mathrm{b}}_\mathrm{R}}\xspace}
\providecommand{\sbone}{\ensuremath{\tilde{\mathrm{b}}_1}\xspace}
\providecommand{\sbtwo}{\ensuremath{\tilde{\mathrm{b}}_2}\xspace}
\providecommand{\sLep}{\ensuremath{\tilde{\mathrm{l}}}\xspace}
\providecommand{\sLepC}{\ensuremath{\tilde{\mathrm{l}}^\mathrm{C}}\xspace}
\providecommand{\sEl}{\ensuremath{\tilde{\mathrm{e}}}\xspace}
\providecommand{\sElC}{\ensuremath{\tilde{\mathrm{e}}^\mathrm{C}}\xspace}
\providecommand{\seL}{\ensuremath{\tilde{\mathrm{e}}_\mathrm{L}}\xspace}
\providecommand{\seR}{\ensuremath{\tilde{\mathrm{e}}_\mathrm{R}}\xspace}
\providecommand{\snL}{\ensuremath{\tilde{\nu}_L}\xspace}
\providecommand{\sMu}{\ensuremath{\tilde{\mu}}\xspace}
\providecommand{\sNu}{\ensuremath{\tilde{\nu}}\xspace}
\providecommand{\sTau}{\ensuremath{\tilde{\tau}}\xspace}
\providecommand{\Glu}{\ensuremath{\mathrm{g}}\xspace}
\providecommand{\sGlu}{\ensuremath{\tilde{\mathrm{g}}}\xspace}
\providecommand{\Wpm}{\ensuremath{\mathrm{W}^{\pm}}\xspace}
\providecommand{\sWpm}{\ensuremath{\tilde{\mathrm{W}}^{\pm}}\xspace}
\providecommand{\Wz}{\ensuremath{\mathrm{W}^{0}}\xspace}
\providecommand{\sWz}{\ensuremath{\tilde{\mathrm{W}}^{0}}\xspace}
\providecommand{\sWino}{\ensuremath{\tilde{\mathrm{W}}}\xspace}
\providecommand{\Bz}{\ensuremath{\mathrm{B}^{0}}\xspace}
\providecommand{\sBz}{\ensuremath{\tilde{\mathrm{B}}^{0}}\xspace}
\providecommand{\sBino}{\ensuremath{\tilde{\mathrm{B}}}\xspace}
\providecommand{\Zz}{\ensuremath{\mathrm{Z}^{0}}\xspace}
\providecommand{\sZino}{\ensuremath{\tilde{\mathrm{Z}}^{0}}\xspace}
\providecommand{\sGam}{\ensuremath{\tilde{\gamma}}\xspace}
\providecommand{\chiz}{\ensuremath{\tilde{\chi}^{0}}\xspace}
\providecommand{\chip}{\ensuremath{\tilde{\chi}^{+}}\xspace}
\providecommand{\chim}{\ensuremath{\tilde{\chi}^{-}}\xspace}
\providecommand{\chipm}{\ensuremath{\tilde{\chi}^{\pm}}\xspace}
\providecommand{\Hone}{\ensuremath{\mathrm{H}_\mathrm{d}}\xspace}
\providecommand{\sHone}{\ensuremath{\tilde{\mathrm{H}}_\mathrm{d}}\xspace}
\providecommand{\Htwo}{\ensuremath{\mathrm{H}_\mathrm{u}}\xspace}
\providecommand{\sHtwo}{\ensuremath{\tilde{\mathrm{H}}_\mathrm{u}}\xspace}
\providecommand{\sHig}{\ensuremath{\tilde{\mathrm{H}}}\xspace}
\providecommand{\sHa}{\ensuremath{\tilde{\mathrm{H}}_\mathrm{a}}\xspace}
\providecommand{\sHb}{\ensuremath{\tilde{\mathrm{H}}_\mathrm{b}}\xspace}
\providecommand{\sHpm}{\ensuremath{\tilde{\mathrm{H}}^{\pm}}\xspace}
\providecommand{\hz}{\ensuremath{\mathrm{h}^{0}}\xspace}
\providecommand{\Hz}{\ensuremath{\mathrm{H}^{0}}\xspace}
\providecommand{\Az}{\ensuremath{\mathrm{A}^{0}}\xspace}
\providecommand{\Hpm}{\ensuremath{\mathrm{H}^{\pm}}\xspace}
\providecommand{\sGra}{\ensuremath{\tilde{\mathrm{G}}}\xspace}
\providecommand{\mtil}{\ensuremath{\tilde{m}}\xspace}
\providecommand{\rpv}{\ensuremath{\rlap{\kern.2em/}R}\xspace}
\providecommand{\LLE}{\ensuremath{LL\bar{E}}\xspace}
\providecommand{\LQD}{\ensuremath{LQ\bar{D}}\xspace}
\providecommand{\UDD}{\ensuremath{\overline{UDD}}\xspace}
\providecommand{\Lam}{\ensuremath{\lambda}\xspace}
\providecommand{\Lamp}{\ensuremath{\lambda'}\xspace}
\providecommand{\Lampp}{\ensuremath{\lambda''}\xspace}
\providecommand{\spinbd}[2]{\ensuremath{\bar{#1}_{\dot{#2}}}\xspace}

\providecommand{\MD}{\ensuremath{{M_\mathrm{D}}}\xspace}
\providecommand{\Mpl}{\ensuremath{{M_\mathrm{Pl}}}\xspace}
\providecommand{\Rinv} {\ensuremath{{R}^{-1}}\xspace} 

%% file: EXO-10-017-authorlist.tex
\textbf{Yerevan Physics Institute,  Yerevan,  Armenia}\\*[0pt]
V.~Khachatryan, A.M.~Sirunyan, A.~Tumasyan
\vskip\cmsinstskip
\textbf{Institut f\"{u}r Hochenergiephysik der OeAW,  Wien,  Austria}\\*[0pt]
W.~Adam, T.~Bergauer, M.~Dragicevic, J.~Er\"{o}, C.~Fabjan, M.~Friedl, R.~Fr\"{u}hwirth, V.M.~Ghete, J.~Hammer\cmsAuthorMark{1}, S.~H\"{a}nsel, C.~Hartl, M.~Hoch, N.~H\"{o}rmann, J.~Hrubec, M.~Jeitler, G.~Kasieczka, W.~Kiesenhofer, M.~Krammer, D.~Liko, I.~Mikulec, M.~Pernicka, H.~Rohringer, R.~Sch\"{o}fbeck, J.~Strauss, A.~Taurok, F.~Teischinger, W.~Waltenberger, G.~Walzel, E.~Widl, C.-E.~Wulz
% ...
\vskip\cmsinstskip
% ...
1:~~Also at CERN, European Organization for Nuclear Research, Geneva, Switzerland\\
% ...


%% file: EXO-10-017_temp.bbl
\providecommand{\href}[2]{#2}\begingroup\raggedright\begin{thebibliography}{10}

\bibitem{dl}
S.~Dimopoulos and G.~Landsberg, ``Black Holes at the {LHC}'', \textit{ Phys.
  Rev. Lett.} \textbf{ 87} (2001) 161602,
  \href{http://www.arXiv.org/abs/hep-ph/0106295}{\texttt{
  arXiv:hep-ph/0106295}}.
  \href{http://dx.doi.org/10.1103/PhysRevLett.87.161602}{\texttt{
  doi:10.1103/PhysRevLett.87.161602}}.

\bibitem{gt}
S.~Giddings and S.~Thomas, ``High-energy colliders as black hole factories: The
  end of short distance physics'', \textit{ Phys. Rev.} \textbf{ D65} (2002)
  056010, \href{http://www.arXiv.org/abs/hep-ph/0106219}{\texttt{
  arXiv:hep-ph/0106219}}.
  \href{http://dx.doi.org/10.1103/PhysRevD.65.056010}{\texttt{
  doi:10.1103/PhysRevD.65.056010}}.

\bibitem{add}
N.~Arkani-Hamed, S.~Dimopoulos, and G.~Dvali, ``The Hierarchy problem and new
  dimensions at a millimeter'', \textit{ Phys.\ Lett.} \textbf{ B429} (1998)
  263. \href{http://dx.doi.org/10.1016/S0370-2693(98)00466-3}{\texttt{
  doi:10.1016/S0370-2693(98)00466-3}}.

\bibitem{add1}
N.~Arkani-Hamed, S.~Dimopoulos, and G.~Dvali, ``Phenomenology, astrophysics and
  cosmology of theories with submillimeter dimensions and {TeV} scale quantum
  gravity'', \textit{ Phys. Rev.} \textbf{ D59} (1999) 086004.
  \href{http://dx.doi.org/10.1103/PhysRevD.59.086004}{\texttt{
  doi:10.1103/PhysRevD.59.086004}}.

\bibitem{PDG}
{ Particle Data Group} Collaboration, ``The Review of Particle Physics'',
  \textit{ J. Phys.} \textbf{ G37} (2010) 075021.
  \href{http://dx.doi.org/10.1088/0954-3899/37/7A/075021}{\texttt{
  doi:10.1088/0954-3899/37/7A/075021}}.

\bibitem{mp}
R.~Myers and M.~Perry, ``Black Holes in Higher Dimensional Space-Times'',
  \textit{ Ann. Phys.} \textbf{ 172} (1986) 304.
  \href{http://dx.doi.org/10.1016/0003-4916(86)90186-7}{\texttt{
  doi:10.1016/0003-4916(86)90186-7}}.

\bibitem{adm}
P.~Argyres, S.~Dimopoulos, and J.~March-Russell, ``Black holes and
  submillimeter dimensions'', \textit{ Phys. Lett.} \textbf{ B441} (1998) 96.
  \href{http://dx.doi.org/10.1016/S0370-2693(98)01184-8}{\texttt{
  doi:10.1016/S0370-2693(98)01184-8}}.

\bibitem{Hawking}
S.~Hawking, ``Particle Creation by Black Holes'', \textit{ Commun. Math. Phys.}
  \textbf{ 43} (1975) 199. \href{http://dx.doi.org/10.1007/BF02345020}{\texttt{
  doi:10.1007/BF02345020}}.

\bibitem{EHM}
R.~Emparan, G.~Horowitz, and R.~Myers, ``{Black holes radiate mainly on the
  brane}'', \textit{ Phys. Rev. Lett.} \textbf{ 85} (2000) 499,
  \href{http://www.arXiv.org/abs/hep-th/0003118}{\texttt{
  arXiv:hep-th/0003118}}.
  \href{http://dx.doi.org/10.1103/PhysRevLett.85.499}{\texttt{
  doi:10.1103/PhysRevLett.85.499}}.

\bibitem{RM}
P.~Meade and L.~Randall, ``Black Holes and Quantum Gravity at the{ LHC}'',
  \textit{ JHEP} \textbf{ 05} (2008) 003,
  \href{http://www.arXiv.org/abs/0708.3017}{\texttt{ arXiv:0708.3017}}.
  \href{http://dx.doi.org/10.1088/1126-6708/2008/05/003}{\texttt{
  doi:10.1088/1126-6708/2008/05/003}}.

\bibitem{Calmet}
X.~Calmet, W.~Gong, and S.~Hsu, ``Colorful quantum black holes at the{ LHC}'',
  \textit{ Phys. Lett.} \textbf{ B668} (2008) 20,
  \href{http://www.arXiv.org/abs/0806.4605}{\texttt{ arXiv:0806.4605}}.
  \href{http://dx.doi.org/10.1016/j.physletb.2008.08.011}{\texttt{
  doi:10.1016/j.physletb.2008.08.011}}.

\bibitem{DG}
D.~Gingrich, ``Quantum black holes with charge, colour, and spin at the{
  LHC}'', \textit{ J. Phys.} \textbf{ G37} (2010) 105108,
  \href{http://www.arXiv.org/abs/0912.0826}{\texttt{ arXiv:0912.0826}}.
  \href{http://dx.doi.org/10.1088/0954-3899/37/10/105008}{\texttt{
  doi:10.1088/0954-3899/37/10/105008}}.

\bibitem{dijets1}
{ CMS} Collaboration, ``Search for Dijet Resonances in 7 {TeV} pp Collisions at
  {CMS}'', \textit{ Phys. Rev. Lett.} \textbf{ 105} (2010) 211801,
  \href{http://www.arXiv.org/abs/1010.0203}{\texttt{ arXiv:1010.0203}}.
  \href{http://dx.doi.org/10.1103/PhysRevLett.105.21180}{\texttt{
  doi:10.1103/PhysRevLett.105.21180}}.

\bibitem{dijets2}
{ CMS} Collaboration, {CMS Collaboration}, ``Search for Quark Compositeness
  with the Dijet Centrality Ratio in pp Collisions at $\sqrt{s}=7$ {TeV}''.
  2010. \href{http://www.arXiv.org/abs/1010.4439}{\texttt{ arXiv:1010.4439}}.

\bibitem{review1}
P.~Kanti, ``Black holes in theories with large extra dimensions: A Review'',
  \textit{ Int. J. Mod. Phys.} \textbf{ A19} (2004) 4899,
  \href{http://www.arXiv.org/abs/hep-ph/0402168}{\texttt{
  arXiv:hep-ph/0402168}}.
  \href{http://dx.doi.org/10.1142/S0217751X04018324}{\texttt{
  doi:10.1142/S0217751X04018324}}.

\bibitem{review2}
G.~Landsberg, ``Black Holes at Future Colliders and Beyond'', \textit{ J.
  Phys.} \textbf{ G32} (2006) R337,
  \href{http://www.arXiv.org/abs/hep-ph/0607297}{\texttt{
  arXiv:hep-ph/0607297}}.
  \href{http://dx.doi.org/10.1088/0954-3899/32/9/R02}{\texttt{
  doi:10.1088/0954-3899/32/9/R02}}.

\bibitem{CMS}
{ CMS} Collaboration, ``The {CMS} experiment at the {CERN} {LHC}'', \textit{
  JINST} \textbf{ 3} (2008) S08004.
  \href{http://dx.doi.org/10.1088/1748-0221/3/08/S08004}{\texttt{
  doi:10.1088/1748-0221/3/08/S08004}}.

\bibitem{lumi}
{ CMS} Collaboration, ``Measurement of {CMS} Luminosity'', \textit{ CMS Physics
  Analysis Summary} \textbf{
  \href{http://cdsweb.cern.ch/record/1279145}{CMS-PAS-EWK-10-004}} (2010).

\bibitem{anti-kt}
M.~Cacciari, G.~Salam, and G.~Soyez, ``The anti-$k_T$ jet clustering
  algorithm'', \textit{ JHEP} \textbf{ 04} (2008) 063.
  \href{http://dx.doi.org/10.1088/1126-6708/2008/04/063}{\texttt{
  doi:10.1088/1126-6708/2008/04/063}}.

\bibitem{JES}
{ CMS} Collaboration, ``Jet Energy Corrections determination at 7 {TeV}'',
  \textit{ CMS Physics Analysis Summary} \textbf{
  \href{http://cdsweb.cern.ch/record/1308178}{CMS-PAS-JME-10-010}} (2010).

\bibitem{MET}
{ CMS} Collaboration, ``Missing Transverse Energy Performance in Minimum-Bias
  and Jet Events from Proton-Proton Collisions at $\sqrt{s}=7$ {TeV}'',
  \textit{ CMS Physics Analysis Summary} \textbf{
  \href{http://cdsweb.cern.ch/record/1279142}{CMS-PAS-JME-10-004}} (2010).

\bibitem{ECAL}
P.~Adzic {et~al.}, ``Energy resolution of the barrel of the {CMS}
  electromagnetic calorimeter'', \textit{ JINST} \textbf{ 2} (2007) P04004.
  \href{http://dx.doi.org/10.1088/1748-0221/2/04/P04004}{\texttt{
  doi:10.1088/1748-0221/2/04/P04004}}.

\bibitem{BlackMax}
D.-C. Dai, G.~Starkman, D.~Stojkovic{ et~al.}, ``BlackMax: A black-hole event
  generator with rotation, recoil, split branes, and brane tension'', \textit{
  Phys. Rev.} \textbf{ D77} (2008) 076007.
  \href{http://dx.doi.org/10.1103/PhysRevD.77.076007}{\texttt{
  doi:10.1103/PhysRevD.77.076007}}.

\bibitem{PYTHIA}
T.~Sj{\"o}strand, S.~Mrenna, and P.~Skands, ``{PYTHIA} 6.4 Physics and
  Manual'', \textit{ JHEP} \textbf{ 05} (2006) 026.
  \href{http://dx.doi.org/10.1088/1126-6708/2006/05/026}{\texttt{
  doi:10.1088/1126-6708/2006/05/026}}.

\bibitem{FastSim}
D.~Orbaker, ``Fast Simulation of the {CMS} Detector'', \textit{ J. Phys. Conf.
  Ser.} \textbf{ 219} (2010) 032053.
  \href{http://dx.doi.org/10.1088/1742-6596/219/3/032053}{\texttt{
  doi:10.1088/1742-6596/219/3/032053}}.

\bibitem{GEANT4}
{ GEANT 4} Collaboration, ``{GEANT4} -- a simulation toolkit'', \textit{ Nucl.
  Instr. and Methods} \textbf{ A506} (2003) 250.
  \href{http://dx.doi.org/10.1016/S0168-9002(03)01368-8}{\texttt{
  doi:10.1016/S0168-9002(03)01368-8}}.

\bibitem{MSTW}
A.~D. Martin, W.~J. Stirling, R.~S. Thorne{ et~al.}, ``Heavy-quark mass
  dependence in global {PDF} analyses and 3- and 4-flavour parton
  distributions'', \textit{ Eur. Phys. J.} \textbf{ C70} (2010) 51,
  \href{http://www.arXiv.org/abs/1007.2624}{\texttt{ arXiv:1007.2624}}.
  \href{http://dx.doi.org/10.1140/epjc/s10052-010-1462-8}{\texttt{
  doi:10.1140/epjc/s10052-010-1462-8}}.

\bibitem{CHARYBDIS}
C.~M. Harris, P.~Richardson, and B.~R. Webber, ``{CHARYBDIS}: A black hole
  event generator'', \textit{ JHEP} \textbf{ 08} (2003) 033,
  \href{http://www.arXiv.org/abs/hep-ph/0307305}{\texttt{
  arXiv:hep-ph/0307305}}.

\bibitem{CHARYBDIS2}
J.~Frost {et~al.}, ``Phenomenology of Production and Decay of Spinning Extra-
  Dimensional Black Holes at Hadron Colliders'', \textit{ JHEP} \textbf{ 10}
  (2009) 014, \href{http://www.arXiv.org/abs/0904.0979}{\texttt{
  arXiv:0904.0979}}.
  \href{http://dx.doi.org/10.1088/1126-6708/2009/10/014}{\texttt{
  doi:10.1088/1126-6708/2009/10/014}}.

\bibitem{MadGraph}
J.~Alwall {et~al.}, ``{MadGraph/MadEvent} v4: The New Web Generation'',
  \textit{ JHEP} \textbf{ 09} (2007) 028.
  \href{http://dx.doi.org/10.1088/1126-6708/2007/09/028}{\texttt{
  doi:10.1088/1126-6708/2007/09/028}}.

\bibitem{CTEQ}
{ CTEQ} Collaboration, ``Implications of {CTEQ} global analysis for collider
  observables'', \textit{ Phys. Rev.} \textbf{ D78} (2008) 013004.
  \href{http://dx.doi.org/10.1103/PhysRevD.78.013004}{\texttt{
  doi:10.1103/PhysRevD.78.013004}}.

\bibitem{ALPGEN}
M.~Mangano, M.~Moretti, F.~Piccinini{ et~al.}, ``{ALPGEN}, a generator for hard
  multiparton processes in hadronic collisions'', \textit{ JHEP} \textbf{ 07}
  (2003) 001, \href{http://www.arXiv.org/abs/hep-ph/0206293}{\texttt{
  arXiv:hep-ph/0206293}}.
  \href{http://dx.doi.org/10.1088/1126-6708/2003/07/001}{\texttt{
  doi:10.1088/1126-6708/2003/07/001}}.

\bibitem{Bayes}
I.~Bertram, G.~Landsberg, J.~Linnemann{ et~al.}, ``{A Recipe for the
  construction of confidence limits}'', technical report, 2000.
\newblock
  \href{http://lss.fnal.gov/archive/test-tm/2000/fermilab-tm-2104.pdf}{FERMILA%
B-TM-2104}.

\bibitem{ttbar}
K.~Agashe, A.~Belyaev, T.~Krupovnickas{ et~al.}, ``{LHC} Signals from Warped
  Extra Dimensions'', \textit{ Phys. Rev.} \textbf{ D77} (2008) 015003.
  \href{http://dx.doi.org/10.1103/PhysRevD.77.015003}{\texttt{
  doi:10.1103/PhysRevD.77.015003}}.

\bibitem{sixjet1}
R.~Chivukula, M.~Golden, and E.~Simmons, ``Six jet signals of highly colored
  fermions'', \textit{ Phys. Lett.} \textbf{ B257} (1991) 403.
  \href{http://dx.doi.org/10.1016/0370-2693(91)91915-I}{\texttt{
  doi:10.1016/0370-2693(91)91915-I}}.

\bibitem{sixjet2}
R.~Chivukula, M.~Golden, and E.~Simmons, ``Multi-jet physics at hadron
  colliders'', \textit{ Nucl. Phys.} \textbf{ B363} (1991) 83.
  \href{http://dx.doi.org/10.1016/0550-3213(91)90235-P}{\texttt{
  doi:10.1016/0550-3213(91)90235-P}}.

\bibitem{coloron}
B.~Dobrescu, K.~Kong, and R.~Mahbubani, ``Massive color-octet bosons and pairs
  of resonances at hadron colliders'', \textit{ Phys. Lett.} \textbf{ B670}
  (2008) 119, \href{http://www.arXiv.org/abs/arXiv:0709.2378v3
  [hep-ph]}{\texttt{ arXiv:arXiv:0709.2378v3 [hep-ph]}}.
  \href{http://dx.doi.org/10.1016/j.physletb.2008.10.048}{\texttt{
  doi:10.1016/j.physletb.2008.10.048}}.

\bibitem{SB}
S.~Dimopoulos and R.~Emparan, ``String balls at the {LHC} and beyond'',
  \textit{ Phys. Lett.} \textbf{ B526} (2002) 393,
  \href{http://www.arXiv.org/abs/hep-ph/0108060}{\texttt{
  arXiv:hep-ph/0108060}}.
  \href{http://dx.doi.org/10.1016/S0370-2693(01)01525-8}{\texttt{
  doi:10.1016/S0370-2693(01)01525-8}}.

\end{thebibliography}\endgroup
